\documentclass[a4paper,11pt]{article}
\pdfoutput=1 

\usepackage{jheppub} 
\usepackage[T1]{fontenc} 
\usepackage{dsfont}
\usepackage{amsthm}
\usepackage{amsmath}
\usepackage{amsfonts}
\usepackage{tensor}
\usepackage{mathtools}
\usepackage{graphicx}
\usepackage{amssymb}
\usepackage{bm}
\usepackage{setspace}
\usepackage{enumitem}
\usepackage{mdwlist}
\usepackage{parskip}
\usepackage{listings}
\usepackage{color}
\usepackage{hyperref}
\usepackage{subcaption}

\usepackage[normalem]{ulem}
\usepackage[dvipsnames]{xcolor}

\newcommand{\ads}{\text{AdS}}

\newcommand{\dif}{\mathrm{d}}
\newcommand{\avg}[1]{\left\langle #1 \right\rangle}

\newcommand{\mathsc}[1]{{\normalfont\textsc{#1}}}

\DeclarePairedDelimiter\abs{\lvert}{\rvert}%

\title{\boldmath Screening of Coulomb interactions in Holography}

\author{E. Mauri,}
\author{H.T.C. Stoof}

\affiliation{Institute for Theoretical Physics, Utrecht University,\\Princetonplein 5, 3584CC Utrecht, The Netherlands}

\emailAdd{e.mauri@uu.nl}
\emailAdd{H.T.C.Stoof@uu.nl}

\abstract{We introduce Coulomb interactions in the holographic description of strongly interacting systems by performing a (current-current) double-trace deformation of the boundary theory.
In the theory dual to a Reissner-Nordstr\"om background, this deformation leads to gapped plasmon modes in the density-density response, as expected from conventional RPA calculations.  
We further show that by introducing a $(d + 1)$-dimensional Coulomb interaction in a boundary theory in $d$ spacetime dimensions, we recover plasmon modes whose dispersion is proportional to $\sqrt{\abs{\bm k}}$, as observed for example in graphene layers.
Moreover, motivated by recent experimental results in layered cuprate high-temperature superconductors, we present a toy model for a layered system consisting of an infinite stack of (spatially) two-dimensional layers that are coupled only by the long-range Coulomb interaction. This leads to low-energy `acoustic plasmons'. 
Finally, we compute the optical conductivity of the deformed theory in $d = 3 + 1$, where a logarithmic correction is present, and we show how this can be related to the conductivity measured in Dirac and Weyl semimetals.
}

\begin{document} 
\maketitle
\flushbottom

\section{Introduction}
\label{sec:intro}
The AdS/CFT correspondence \cite{Maldacena1998,Gubser1998,Witten1998} has become an important tool for studying strongly coupled quantum field theories, and, in the last decade, it found an ever-increasing range of applications into the realm of condensed-matter physics \cite{Herzog2009,Hartnoll2009,Zaanen2015,Hartnoll2016}. Many recent experiments demonstrate that in strongly interacting systems as, for example, the cuprates high-temperature superconductors, the observed behavior can deviate quite drastically from well known condensed-matter results based on weakly coupled theories that admit a quasiparticles description. In particular, these materials exhibit a strange-metal phase for temperatures higher than the superconducting critical temperature, where quasiparticle excitations appear to be absent (see e.g.\ ref.\ \cite{Hartnoll2016}). The promising aspect of the gauge/gravity duality is that it allows us to study systems without quasiparticle excitations. It can be used to compute thermodynamic quantities and response functions with relatively little effort by mapping strongly interacting quantum field theories on the boundary of an asymptotically AdS spacetime, to classical gravity in the curved bulk spacetime. Trying to tune holographic models such that they reproduce the properties of laboratory condensed-matter systems as closely as possible is, therefore, one important goal of ongoing research in holographic applications to condensed matter.

A common feature in conventional holographic calculations is that they describe neutral systems, where the low-energy hydrodynamic excitations in the longitudinal channel contain sound modes. However, it is well established that in metals the  density fluctuations of the charged electrons give rise to a different type of collective excitations known as plasmons \cite{stoof2008ultracold}. In order to have more realistic models of strongly interacting metals and superconductors, it is thus important to modify the holographic theory to include the effect of the long-range Coulomb interaction that turns the linear sound modes into plasmon modes. 

In textbook condensed-matter response calculations the electron-electron interaction is introduced in the description through a self-consisted field method known as the random-phase approximation (RPA) \cite{PhysRev.82.625,PhysRev.85.338,PhysRev.92.609,PhysRev.92.626}, where the induced charge without Coulomb interaction is replaced with the screened charge. In a translational invariant system the density-density response function $\chi(\omega, \bm k)$ then becomes, in terms of the non-interacting response function $\chi_0(\omega, \bm k)$, equal to
\begin{align}
  \chi(\omega, \bm k) =  \frac{\chi_0(\omega, \bm k)}{1 - V(\omega, \bm k)\chi_0(\omega, \bm k)} \equiv \frac{\chi_0(\omega, \bm k)}{\epsilon(\omega, \bm k)} \text{ ,}
\end{align}
where $\epsilon(\omega, \bm k)$ is the dynamical dielectric function, and $V(\omega, \bm k)$ the Fourier transform of the Coulomb potential. In non-relativistic systems, where the Fermi velocity is much smaller than the speed of light, the latter is only a function of the magnitude of the momentum $\abs{\bm k}$. In particular, for 2D and 3D materials, where here we refer to the spatial dimension of a system living in the usual $(3 + 1)$-dimensional spacetime, we have that
\begin{align}\label{eq:rpa_cond_matt}
  \begin{cases}
    V(\omega, \bm k) = \frac{e^2}{2 \epsilon_0 \abs{\bm k}} , & 2\text{D} \\
    V(\omega, \bm k) = \frac{e^2}{\epsilon_0 \bm k^2} , & 3 \text{D} \text{ ,}
  \end{cases} 
\end{align}
with $-e$ the charge of the electron, and $\epsilon_0$ the dielectric constant.

In this paper, we develop a generic procedure to include the above RPA correction, and hence the effect of Coulomb interactions, into holographic models at nonzero density. A first step towards studying plasmon modes in holography was recently made in refs.\ \cite{Gran2017,Aronsson2018,Gran2018,Gran2018a}, where it was shown that plasmon quasinormal-modes can be studied by imposing a new type of boundary conditions on the bulk linearized equations of motion. Here we explain how these boundary conditions naturally arise by introducing a (current-current) double-trace deformation of the boundary theory, and how this gives rise to an RPA-like form of the response function. Using also the example of the Reissner-Nordstr\"om metal, we numerically study the full frequency and momentum dependence of the spectral functions and show how the prescribed procedure turns the low-energy sound modes into gapped plasmon modes. 

In the particular case of a $(3+1)$-dimensional theory, we also show that by introducing long-range Coulomb interactions in the holographic theory, we obtain an anomalous cutoff-dependent logarithmic behavior in the real part of the optical conductivity, analogous to the one observed in the conductivity of Dirac and Weyl semimetals \cite{Roy2017}. 

In $(2+1)$-dimensions, we adapt our procedure to correctly describe a system where electrons are constrained to two spatial dimensions, but where the Coulomb potential exists in three spatial dimensions. This allows us to obtain a low-energy dispersion relation $\omega \propto \sqrt{\abs{\bm k}}$ as expected in charged (spatially) two-dimensional system as, for example, graphene \cite{Lucas2018a,Lucas2018}. Furthermore, since experiments on $(2+1)$-dimensional materials are often performed on a multi-layer system, we propose a toy model describing an infinite stack of $(2+1)$-dimensional layers, where the dynamics in each layer is determined holographically through a dual Reissner-Nordstr\"om theory and where the coupling between the layers is given only by the long-range Coulomb interaction. A recent example of the interest in these multi-layered systems is given in ref.\ \cite{Hepting2018}, where the spectral function for a layered copper-oxide high-temperature superconductor has been measured. These experimental results suggest that it is indeed the Coulomb interaction that governs the coupling between different layers, validating the assumptions of our toy model, and hinting at the fact that low-energy charge fluctuations might be very relevant for the description of the dynamics of high-$T_c$ superconductors. We, therefore, show how our toy model qualitatively reproduces the observed spectral functions, with low-energy linear excitations, often referred to as `acoustic plasmons', believed to possibly play an important role in the mechanism of high-temperature superconductivity \cite{Ishii1993,Kresin1988,article}. 
  
The structure of the paper is as follow. In section \ref{sec:reissner_nord} we briefly summarize the Reissner-Nordstr\"om model and its spectral functions containing the sound modes. In  section \ref{sec:double_trace} we explain how to deform the theory to describe a charged system with plasmon excitations and show how the density-density response function obtains the RPA-like form. In section \ref{sec:2dplasmon} we present the numerical results for the spectral functions with plasmon excitations in a $(2+1)$-dimensional theory and propose a toy model for a layered system in $3+1$ dimensions. Finally, section \ref{sec:3dplasmon} contains the results for the optical conductivity for a $(3+1)$-dimensional system with Coulomb interactions.

\section{Reissner-Nordstr\"om black brane}\label{sec:reissner_nord}
  In this section we introduce the Reissner-Nordstr\"om model that we use throughout the paper as a concrete example of the proposed procedure, that we introduce in detail in section \ref{sec:double_trace}. The dynamics of this holographic model has been thoroughly studied in refs.\ \cite{Edalati2010a,Edalati2010, Davison2011, Davison2013}. In order to be able to compare directly with our later results, however, we briefly review its dynamical properties by computing the spectral functions. We then recover that the low-energy excitations of the theory in the longitudinal sector are linear sound modes and a diffusive mode. 

  Throughout the paper, we call $d$ the spacetime dimension of the boundary theory, dual to a gravitational $(d+1)$-dimensional bulk theory and we use the mostly-plus metric. We denote with $r$ the coordinate of the extra spatial dimension of the bulk theory, with the boundary at $r\to \infty$ unless explicitly stated otherwise. When Fourier transforming in the direction orthogonal to $r$, we denote with $k = (\omega/c, \bm k)$ the $d$-dimensional momentum. We use, with a slight abuse of notation, greek lower-case tensor indices for both the $(d + 1)$-dimensional bulk and the $d$-dimensional boundary, as the range of the indices is clear from the context. Moreover, in computing dynamical quantities, we always use a gauge in which all the $r$-components of the field fluctuations are zero, so that no confusion should arise.

  \subsection{Gravity action}
    The simplest holographic model with a nonzero-density boundary theory is the Reissner-Nordstr\"om model, described by a bulk Einstein-Maxwell action, that in SI units reads
    \begin{align}\label{eq:einstein-maxwell_action}
    S_{EM} = \int \mathrm d^{d+1} x\, \sqrt{-g}\left(\frac{c^3}{16\pi G}(R - 2\Lambda)- \frac{1}{4 \lambda c}F_{\mu\nu}F^{\mu\nu}\right) \text{ ,}
    \end{align}
    where $x = (r,c t, \bm x)$, with $r$ the bulk coordinate, $G$ is Newton's gravitational constant in $d + 1$ dimensions, and $\lambda$ is a coupling constant with dimension $\left[\lambda\right] = \mathrm{m}^{d-3}\, \mathrm{kg}$, such that $\left[A_\mu\right] = \mathrm{m}\,\mathrm{kg}\,\mathrm{s}^{-1}$. Defining $A_t \equiv c A_0$,  we then see that  $\left[A_t\right] = \mathrm{m}^2\,\mathrm{kg}\,\mathrm{s}^{-2} = \left[\mu\right]$, i.e., the bulk field $A_t$ has the dimensions of a chemical potential. 
    Since we want to describe a homogeneous and isotropic boundary theory, i.e., the chemical potential and the energy-momentum tensor do not depend on the spacetime coordinates and there is no preferred direction, we look for a solution of the form $A_\mu = (0, A_0(r), \bm 0)$ and $g_{\mu\nu} = g_{\mu\nu}(r)$. 
    
    For notational convenience, we first rewrite the action in terms of dimensionless fields and coordinates, by redefining the fields as
    \begin{align}
      \begin{split}
        \tilde A_t =& \frac{L}{2 r_0} \sqrt{\frac{16\pi G}{\lambda c^4}} \frac{A_t}{c} \text{ ,}\\
        \tilde A_x =& \frac{L}{2 r_0} \sqrt{\frac{16\pi G}{\lambda c^4}} A_x \text{ ,}\\
        \tilde A_r =& \frac{r_0}{2 L} \sqrt{\frac{16\pi G}{\lambda c^4}} A_r \text{ ,}\\
        \tilde g_{\mu\nu} =& \frac{L^2}{r_0^2} g_{\mu\nu} \text{ ,}\\
        (\tilde t, \tilde {\bm x}, \tilde r) =& \left(\frac{c t r_0}{L^2}, \frac{x r_0}{L^2}, \frac{r}{r_0}\right) \text{ ,}
      \end{split}
    \end{align}
    with $L^2 = -d(d - 1)/2 \Lambda$ the AdS radius squared and $r_0$ denotes the position of the black-brane (outer) horizon. 
    Notice that this rescaling thus fixes the horizon at $\tilde r_0 = 1$. However, using the scaling symmetry of the action
    \begin{align}
      \tilde r \rightarrow a \tilde r \text{, } (\tilde t, \tilde {\bm x})  \rightarrow (\tilde t, \tilde {\bm x})/a \text{, } \tilde A_\mu \rightarrow a \tilde A_\mu\text{, }
    \end{align}
    we can obtain any solution from the solution with $\tilde r_0 = 1$.
    From now on we only use dimensionless variables and omit the tildes.  
    In the end, the theory we are going to study, including boundary terms, is described by the action:
    \begin{align}\label{eq:dimless_einstein-maxwell_action}
      \begin{split}
      S/\hbar = \frac{c^3 L^{d -1}}{4 \pi\hbar G} \bigg(&\frac{1}{4}\int_{\mathcal M} \mathrm d^{d+1} x\, \sqrt{-g}\left[R - d( d - 1) - F_{\mu\nu}F^{\mu\nu}\right] + \frac{1}{2}\int_{\partial \mathcal M} \dif^d x \sqrt{-\gamma} K\\ 
      &+ \int_{\partial \mathcal M} \dif^d x \mathcal L^{(d)}_{c.t.}\bigg)\text{ ,}
      \end{split}
    \end{align}
    where now every field and coordinate is dimensionless, as is the prefactor in front of the integral that for now we take to be equal to one, by an appropriate scaling of the action, but we briefly come back to this prefactor later on. The second term is the Gibbons-Hawking-York boundary term, with $\gamma_{\mu\nu} \equiv g_{\mu\nu} - n_\mu n_\nu$ the induced metric on the boundary, $n_\mu$ the unit vector normal to the boundary, and the determinant $\gamma$ of $\gamma_{\mu\nu}$ is taken only in the directions orthogonal to $n^\mu$. Finally, $K$ is the trace of the second fundamental form $K = \gamma^{\mu\nu} \nabla_\mu n_\nu$. This boundary term is necessary to have a well-defined Dirichlet problem on the boundary \cite{York1972,Gibbons1977,Hawking1996}. In the case at hand (see below eq.\ \eqref{eq:sol_dimless} where the asymptotically AdS spacetime is a foliation of flat Minkowski spacetime) we have $n_\nu = \sqrt{g_{rr}} \delta^r_\mu$ and $K$ reduces to $K = -\sqrt{g_{rr}} \gamma^{\mu\nu}\Gamma^r_{\mu\nu}$, with $\Gamma^r_{\mu\nu}$ the bulk Christoffel symbols. The last term in the action contains the counterterms necessary to regulate the divergences in the theory, and its explicit form depends on the dimension $d$ of the boundary spacetime. We present later the counterterms in the case of $d = 2 + 1$ and $d = 3 + 1$. For a detailed treatment see, for instance, ref.\ \cite{DeHaro2001}.

    The classical equations of motion derived from the above action are solved by the Reissner-Nordstr\"om background, describing an asymptotically AdS charged black brane, that in the units we introduced takes the form
    \begin{align}
      \begin{split}\label{eq:sol_dimless}
      \dif s^2 =& -f(r) \mathrm d t^2 + \frac{1}{f(r)} \dif r^2 + r^2 \dif\bm x^2 \text{ ,}\\
        A_t(r) =& \mu \left(1 - {r}^{2-d}\right) \text{ ,}\\
        f(r) =& r^2\left[1 - \left(1 +  \frac{2(d-2)}{d-1}\mu^2 \right)r^{- d} + \frac{2(d-2)}{d-1}\mu^2r^{2(1 - d)}\right] \text{ ,}
      \end{split}
    \end{align}
    where we defined $\lim_{r \to \infty} A_t(r) = \mu \equiv A_t'(1)/(d - 2)$ for $d > 2$, that, according to the holographic dictionary, we interpret as the chemical potential of the boundary theory. Here and below, the prime denotes derivatives with respect to the radial coordinate. From the solution of $f(r)$, we can see that the mass $M$ and charge $Q$ of the black brane are expressed in terms of $\mu$ as
    \begin{align}
      Q =& \sqrt{\frac{2(d-2)}{d-1}}\mu \text{ ,}\\
      M =& 1 + \frac{2(d-2)}{d-1}\mu^2 = 1 + Q^2 \text{ .}
    \end{align}
    The solution is then completely characterized by one dimensionless parameter, that we take to be $T/\mu$, with the black-brane temperature given by
    \begin{align}\label{eq:bhtemperature}
      T(\mu) = \frac{f'(1)}{4\pi} = \frac{d - \frac{2(d - 2)^2}{d - 1}\mu^2}{4 \pi}\text{ .}
  \end{align}
  The equilibrium properties of the theory are well known (see e.g.\ ref.\ \cite{Zaanen2015}) and can be derived from the GKPW rule after regularizing the action at the cutoff scale $r = r_{\mathsc{uv}}$ by inserting the boundary counterterm 
  \begin{align}\label{eq:counterterm}
        S_{c.t.} = - \frac{(d - 1)}{2}\int_{r = r_{\mathsc{uv}}} \dif^d x \sqrt{-\gamma} \text{ .}   
  \end{align}
  We give here for later convenience the nonzero components of the equilibrium expectation values in the chosen units, that for $d > 2$,  are:
  \begin{align}\label{eq:equilibrium_pressure_density}
    \delta^{ij}\avg{P}\equiv& \avg{T^{ij}} = \lim_{r_\mathsc{uv} \to \infty} 2 r_\mathsc{uv}^2 \frac{\delta S^{cl} \vert_{r = r_\mathsc{uv}}}{\delta g_{ij}^{(0)}} = \frac{1}{4}\delta^{ij}\left(1 + \frac{2(d - 2)}{d -1} \mu^2\right) = \frac{1}{4}\delta^{ij} M \text{ ,}\\ \label{eq:equilibrium_energy_density}
        \avg{\epsilon}\equiv&\avg{T^{00}} = \lim_{r_\mathsc{uv} \to \infty} 2 r_\mathsc{uv}^2 \frac{\delta S^{cl} \vert_{r = r_\mathsc{uv}}}{\delta g_{00}^{(0)}} = \frac{(d - 1)}{4}\left(1 + \frac{2(d - 2)}{d -1} \mu^2\right) = \frac{(d -1)}{4}M \text{ ,}\\ \label{eq:equilibrium_density}
        \avg{\rho} \equiv& \avg{J^0} = \lim_{r_\mathsc{uv} \to \infty} \frac{\delta S^{cl} \vert_{r = r_\mathsc{uv}}}{\delta A_{0}^{(0)}} =  (d - 2)\mu \text{ .}
  \end{align} 
  Here $S^{cl}$ is a reminder that the action is evaluated at the classical solution and the superscript in $\delta g_{\mu\nu}^{(0)}$ and $\delta A_\mu^{(0)}$ means that we are deriving with respect to the leading-order term in the asymptotic expansion of the field for large $r$. We see here that, in order to compare holographic results to experiments, we should choose, in this bottom-up approach, the prefactor in front of the action in eq.\ \eqref{eq:dimless_einstein-maxwell_action}, that we conveniently scaled away, such to tune the thermodynamic quantities to values close to the experimental ones.

  \subsection{Fluctuations and response functions}
    The background solution of the gravity action in eq.\ \eqref{eq:dimless_einstein-maxwell_action} defines the equilibrium properties of the boundary theory. In order to study the response of the field theory to small perturbations, we have to consider fluctuations of the classical fields on top of the background solution:
    \begin{align}
      \begin{split}
      A_\mu \rightarrow& A_\mu + a_\mu\\
      g_{\mu\nu} \rightarrow& g_{\mu\nu} + h_{\mu\nu}\text{ .}
      \end{split}
    \end{align}
    Exploiting rotational invariance to fix the momentum along the $x$ direction, so that $k = (\omega, \pm\abs{\bm k}, 0, \dots, 0)$, the fluctuations decouple according to their parity under $O(d - 2)$ acting on $x^2, \dots, x^{d-1}$ \cite{Kovtun2005a}. In particular, we are interested in the longitudinal channel associated with spin 0, as it is the one containing the density-density response function that in $d = 3 + 1$ and $d = 2 + 1$ are described by the components
    \begin{align}
      \delta \bm \Phi \equiv (a_t, a_x, a_r, h_{xt}, h_{tt}, h_{xx}, h_{yy} = h_{zz}, h_{rr}, h_{tr}, h_{xr})\text{ ,}
    \end{align}
    and will ultimately lead to hydrodynamic diffusion and sound modes.
    It is convenient to work in a gauge where $h_{r\mu} = 0$, as well as $A_r = a_r = 0$. The set of coupled fluctuations then becomes
    \begin{align}
      \delta \bm \Phi = (a_t, a_x, h_{xt}, h_{tt}, h_{xx}, h_{yy} = h_{zz})\text{ .}
    \end{align}
    Solving the coupled set of linearized equations of motion for these fluctuations, with infalling boundary conditions at the black-brane horizon, allows us, after eventual renormalization of boundary divergences (see ref.\ \cite{Skenderis2002} for a detailed treatment), to compute the retarded Green's functions of the dual boundary theory. To do so, we follow the procedure in ref.\ \cite{Kaminski2010}. A caveat is that to extract the matrix of Green's functions for a set of $M$ coupled fields we need $M$ independent solutions. In the case at hand, however, our choice to work in a gauge where all the radial components are zero does not completely fix the gauge freedom of $a_\mu$ and $h_{\mu\nu}$, and we are left with four gauge degrees of freedom. This implies that we can only find two independent solutions of the linearized equations of motion, corresponding to the two physical degrees of freedom. The remaining solutions needed to compute the full Green's function matrix are pure gauge solutions that can be constructed analytically. In appendix \ref{app:gauge_solutions}, we show how to construct these gauge solutions and we provide their explicit form for the theory at hand.
    From the retarded Green's function $G^R(\omega, \bm k)$ we can obtain the spectral function defined as 
    \begin{align}
      A(\omega, \bm k) \equiv -\frac{2}{\pi} \text{Im}[G^R(\omega, \bm k)] \text{ .}
    \end{align}
    The spectral function is an interesting quantity as it tells us the rate of work done on the system by a small external perturbation at a given frequency \cite{stoof2008ultracold}. Sharp peaks in the spectral function, that correspond to poles in the Green's function, denote long-lived collective modes, with a lifetime determined by the width of the peak. 

\begin{figure}
      \centering 
      \includegraphics[width=.49\textwidth]{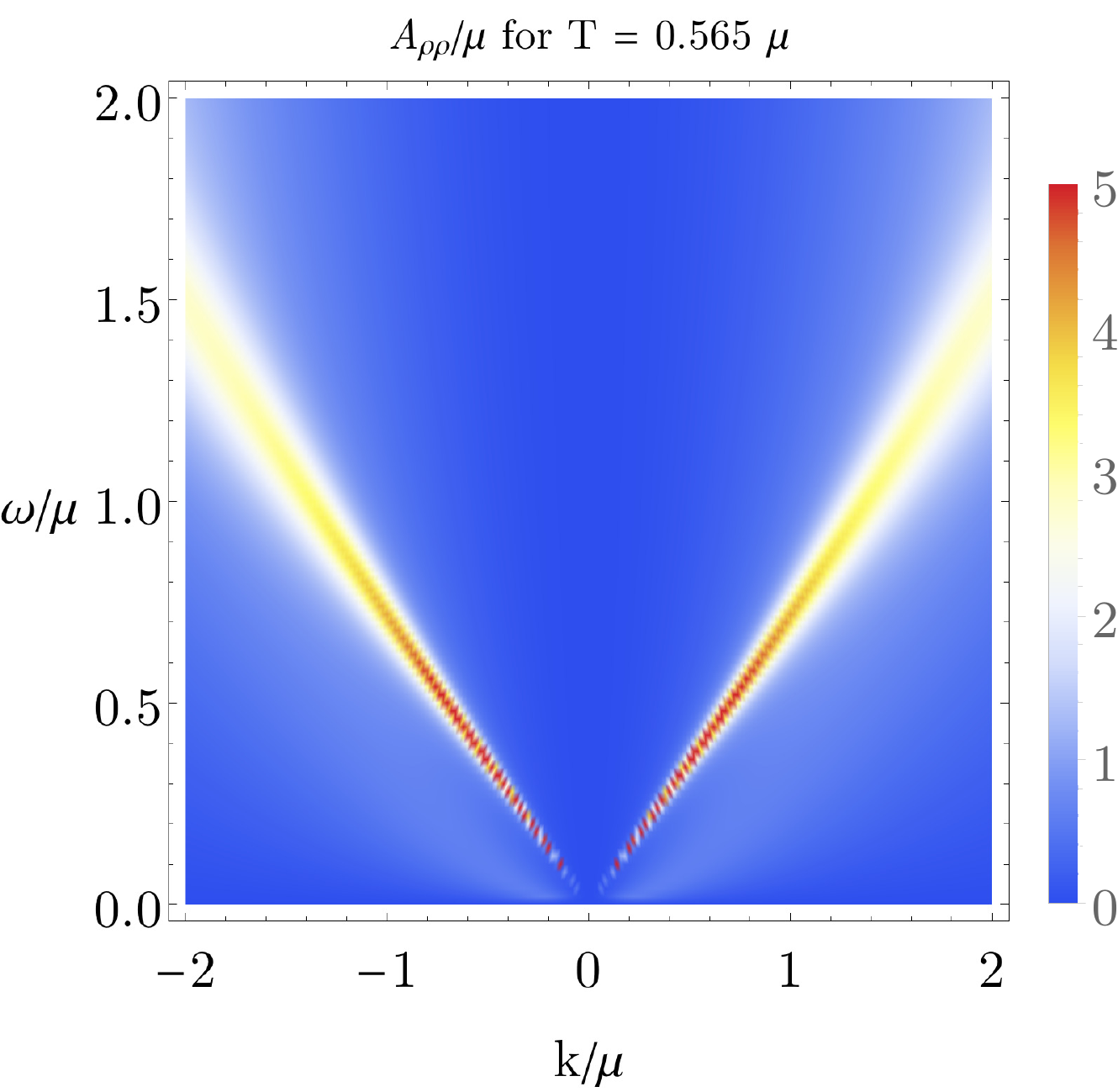}
      \hfill
      \includegraphics[width=.49\textwidth]{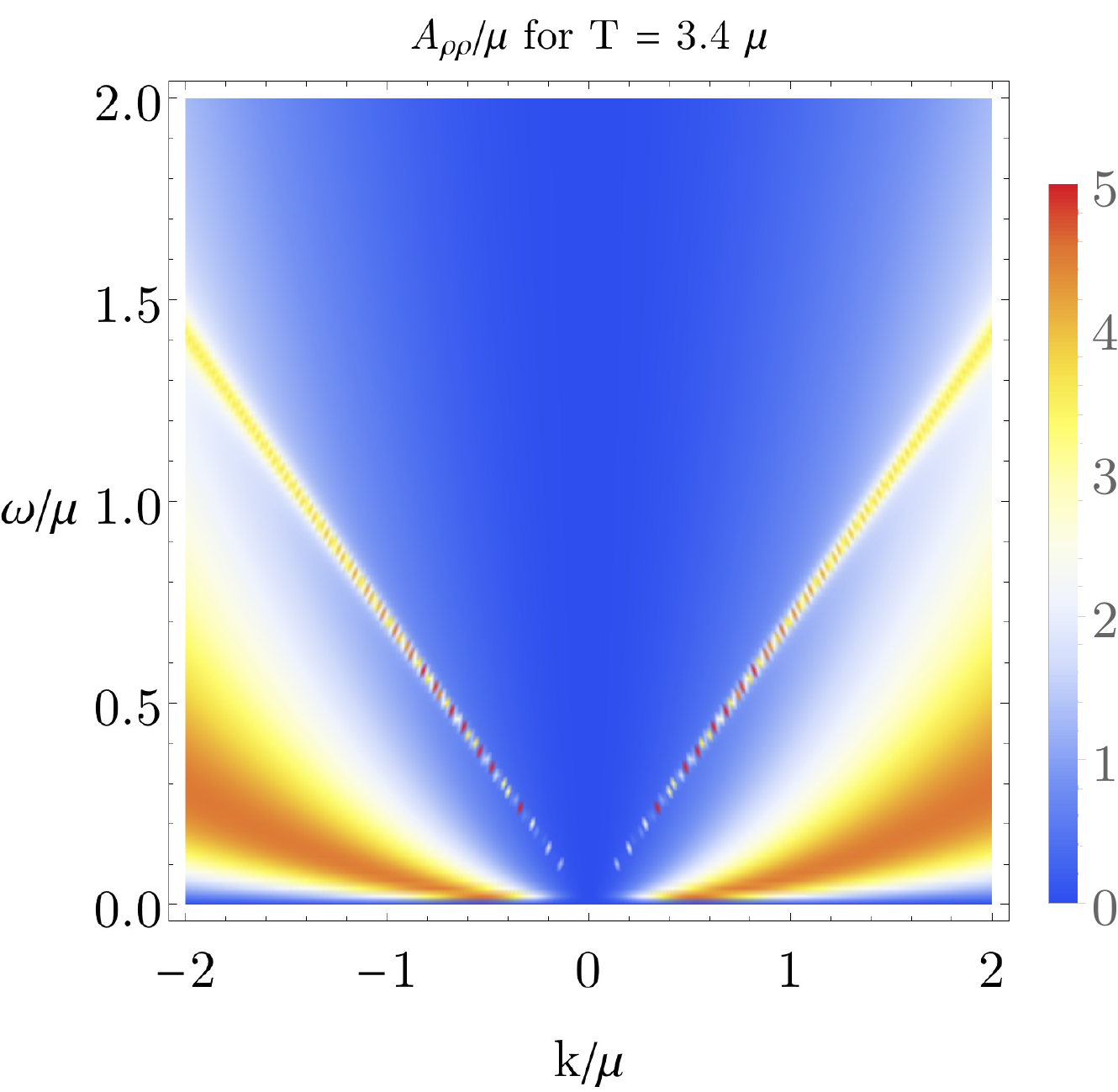}
      \caption{\label{fig:RNdendenSpec} The density spectral function for two different values of the temperature in $2 + 1$ dimensions. Notice that in addition to the linear sound modes with a low-energy behavior of the dispersion given by $\omega = \pm \abs{\bm k}/\sqrt{2}$, we can clearly observe the diffusive mode, that becomes more dominant as we increase $T/\mu$.}
    \end{figure}

\begin{figure}
    \centering 
    \includegraphics[width=.49\textwidth]{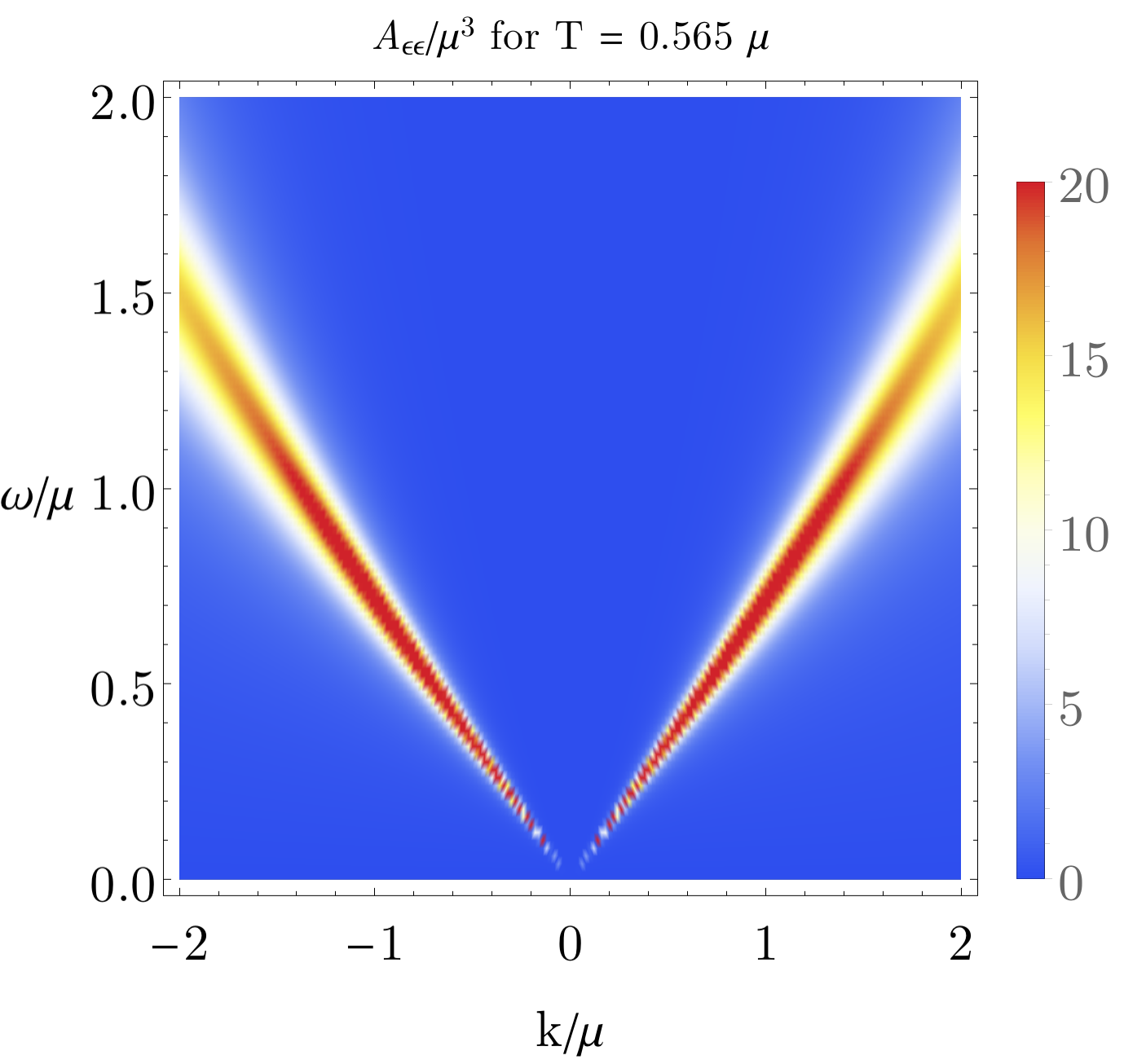}
    \hfill
    \includegraphics[width=.49\textwidth]{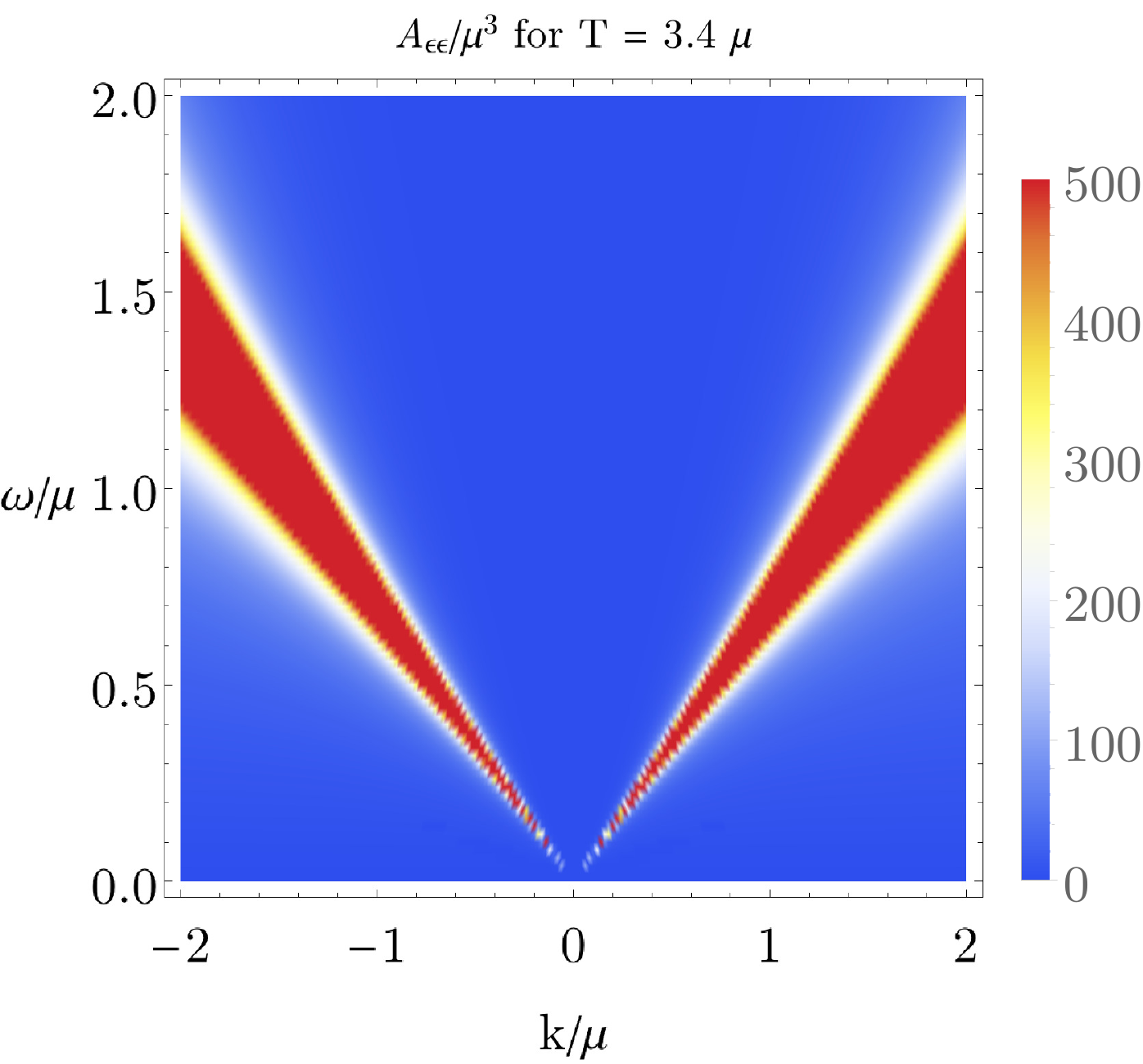}
    \caption{\label{fig:RNendenSpec} The energy-density spectral function for two different values of the temperature in $2 + 1$ dimensions. The sound modes are clearly visible and we can notice that the associated peaks become higher for a higher value of the temperature.}
  \end{figure}

    In figures \ref{fig:RNdendenSpec} and \ref{fig:RNendenSpec} we present the spectral functions of a $(2+1)$-dimensional system for the density response $A_{\rho\rho}$ as well as for the energy-density response $A_{\epsilon\epsilon}$ for two different values of the temperature. Here, as in the rest of the paper, we work in the grand-canonical ensemble, hence all the plots are rescaled to a solution with unit chemical potential. Since we are describing a strongly interacting system, we expect from hydrodynamics considerations (see e.g.\ ref.\ \cite{Kovtun2012}) to find two linear sound modes due to energy-momentum conservation. As the poles are shared by all Green's functions, although the residues differ, in both spectral functions the low-energy linear sound modes $\omega = \pm v_s \abs{\bm k}$ are evident. In particular, up to the numerical precision obtained, the speed of sound satisfies the relation $v_s = 1/\sqrt{d - 1}$ in the low-energy regime \cite{Policastro2002, Davison2011}. These hydrodynamic sound modes, known as first sound, are present even in the low-temperature limit as long as $\omega, k \ll \mu$, with the nonzero chemical potential setting an effective hydrodynamic scale \cite{Edalati2010, Davison2011, Davison2013}. This is similar to an ultra-cold Fermi gas at unitarity where, in the limit $T\to 0$, hydrodynamic first sound exists at long wavelengths in a range set by the Fermi energy, and where this first sound mode crosses over to collisionless zeroth sound at shorter wavelengths. In the density spectral function we observe, in addition to the sound modes, the expected diffusive mode due to charge conservation that corresponds to the broad quadratic band. Here the linear modes become less sharply peaked when the temperature is raised as part of the spectral weight shifts into the diffusive mode. Note that the diffusive mode is absent in $A_{\epsilon\epsilon}$, because in that mode the charge fluctuations do not induce a velocity field $\bm v(t, \bm x)$ in the fluid, hence also no energy transport. A thorough analysis of these modes has been presented for $(2+1)$-dimensional systems in ref.\ \cite{Davison2011}. In the next section, we turn to the problem of deforming the holographic theory in order to describe a charged quantum field theory with plasmon excitations.

  \section{Plasmon modes in holographic theories}\label{sec:double_trace}
    In this section we explain how to introduce Coulomb interactions in the boundary quantum field theory, and show how this gives rise to an RPA-like form of the response functions  that turns the linear modes observed in the Reissner-Nordstr\"om spectral function into long-lived plasmon excitations. 

    The holographic theory presented in the previous section is dual to a charge-neutral boundary theory, that is, the Maxwell field in the bulk acts only as a source for the current operator $J^\mu$ in the boundary theory, and there is no dynamical photon \cite{Hartnoll2008}. However, in condensed matter we are often interested in describing charged systems, i.e., systems with a long-range Coulomb interaction. We, therefore, need to modify the boundary theory to add dynamical photons coupled to the conserved current $\avg{J^\mu}$ obtained from holography in the large-$N$ limit, which are described by the Maxwell action
    \begin{align}\label{eq:qed_action}
      S_\mathsc{m} = \int \dif^d x \left(-\frac{1}{4 \mu_0 e^2} F_{\mu \nu} F^{\mu \nu} + A_\mu \avg{J^\mu} \right) \text{ ,}
    \end{align}
    with $\mu_0$ the magnetic permeability and $-e$ the electron charge. Note that we also used the fact that our boundary theory is defined on flat Minkowski spacetime, so that $\sqrt{-\eta} = 1$.
    Integrating out the $A_\mu$ field from eq.\ \eqref{eq:qed_action}, using the standard $\xi$ procedure to account for the gauge freedom, we obtain in momentum space (see for example ref.\ \cite{matthewschwartz2014})
    \begin{align}\label{eq:maxwell_qed}
      S = -\frac{1}{2}\int \frac{\dif^d k}{(2 \pi)^d} \mu_0 e^2 \avg{J^\mu(k)}\left[\frac{\eta_{\mu\nu} - (1 - \xi)k_\mu k_\nu/k^2}{k^2}\right] \avg{J^\nu(- k)} \text{ .}
    \end{align}
    Notice that eq.\ \eqref{eq:maxwell_qed} assumes the form of a double-trace deformation of the boundary theory for the current operator. 
    
    As first described in refs.\ \cite{Witten2001, Muck2002}, a double-trace deformation in the large-$N$ limit can be incorporated in the holographic dual simply as a change in the boundary conditions for the corresponding fields. In the gauge/gravity duality without multi-trace deformation, we can obtain the response to small fluctuations from the asymptotic behavior of the field fluctuations. For example in the case of the Maxwell-field fluctuations we have
    \begin{align}\label{eq:maxwell_expansion}
      a_\mu = a_\mu^{(0)} + \dots + \frac{\eta_{\mu\nu}}{(d - 2)} \delta\avg{J^\nu}r^{-d + 2} + \mathcal O(r^{-d -2}) \text{ ,}
    \end{align} 
    and according to the holographic dictionary, we interpret the leading-order coefficient $a_\mu^{(0)}$ as the source $a^\mathsc{s}_\mu$ of the fluctuations $\delta\avg{J^\mu}$. Following the prescription in ref.\ \cite{Witten2001}, in order to include a multi-trace deformation of the response $\int \dif^d x\, \delta\avg{J^\mu} \mathcal M_{\mu\nu} \delta\avg{J^\nu} \equiv W(\delta\avg{J^\mu})$, we still interpret the subleading term in equation \eqref{eq:maxwell_expansion} as the response to the source fluctuations, but we impose the boundary condition 
    \begin{align}\label{eq:neumman_bdy_0}
      \lim_{r \to \infty} a_\mu = a^\mathsc{s}_\mu + \frac{\delta W}{\delta \avg{J^\mu}} \text{ .}
    \end{align}
    The reason for this boundary condition can be easily understood by looking at the boundary behavior of the holographic theory. The boundary contribution from the Maxwell-field fluctuations in the renormalized holographic action, eq.\ \eqref{eq:dimless_einstein-maxwell_action}, evaluated on-shell is given, at second order in fluctuations, by
    \begin{align}
      \delta S^{(2)} = \frac{1}{2}\int_{r = \infty} \frac{\dif^d k}{(2 \pi)^d} a_\mu^\mathsc{s}(k)  \delta\avg{J^\mu(- k)} \text{ ,}
    \end{align}
    with
    \begin{align}
      \lim_{r \to \infty} a_\mu = a^{(0)}_\mu = a^\mathsc{s}_\mu \text{ .}
    \end{align}
    By introducing a boundary contribution of the form of eq.\ \eqref{eq:qed_action} at infinity, that, in the saddle point approximation of the large-$N$ limit, we write in dimensionless units as
    \begin{align}\label{eq:maxwell_boundary}
      \begin{split}
      S_{d.t.} =& \int \dif^d x \left(-\frac{1}{4 \alpha^2} \left[F_{\mu \nu} F^{\mu \nu} + \frac{2}{\xi}\left(\partial_\mu A^\mu\right)^2\right] + A_\mu \avg{J^\mu} \right)\\
      =& -\frac{1}{2}\int_{r = \infty} \frac{\mathrm d^d  k}{(2 \pi)^d} \alpha^2 \avg{J^\mu( k)}\left[\frac{\eta_{\mu\nu} - (1 - \xi)k_\mu k_\nu/k^2}{k^2}\right] \avg{J^\nu(-  k)} \text{ ,}
      \end{split}
    \end{align}
    with $\alpha^2$ a dimensionless constant that depends on the charge of the system, its permittivity and on the prefactor of the action in eq.\ \eqref{eq:dimless_einstein-maxwell_action}, the boundary term for the fluctuations now becomes
    \begin{align}
      \delta S^{(2)} = \frac{1}{2}\int_{r = \infty} \frac{\dif^d k}{(2 \pi)^d} \left[ a_\mu^s(k) - \alpha^2 \left(\frac{\eta_{\mu\nu} - (1 - \xi)k_\mu k_\nu/k_\mu k^\mu}{k_\mu k^\mu}\right)\delta\avg{J^\nu(k)}\right] \delta\avg{J^\mu(- k)} \text{ .}
    \end{align}
    Since the double-trace deformation does not modify the bulk theory, we see that this is equivalent to the Reissner-Nordstr\"om theory described above, where we simply modify the boundary conditions
    \begin{align}\label{eq:neumman_bdy}
      \lim_{r \to \infty} a_\mu = a_\mu^{(0)} \equiv  a^\mathsc{s}_\mu - \alpha^2 \left(\frac{\eta_{\mu\nu} - (1 - \xi)k_\mu k_\nu/k^2}{k^2}\right)\delta\avg{J^\nu} \text{ ,}
    \end{align}
    that are of the form of eq.\ \eqref{eq:neumman_bdy_0}.\
    Notice that these boundary conditions are equivalent to the ones used in ref.\ \cite{Aronsson2018} to study plasmonic quasi-normal modes in a $d = 2 + 1$ holographic theory, but that were obtained differently by looking for zeros of the dielectric function. 

    The effect of the deformation in eq.\ \eqref{eq:maxwell_boundary} on the response functions is easily derived, and in practice there is no need to solve the boundary value problem anew (e.g. by shooting). In fact, 
    ignoring for the moment the gravitational part, as it does not influence the procedure for the density-density and current-current spectral functions we are interested in, we have from linear-response theory without the double-trace deformation
    \begin{align}\label{eq:linear_response}
      \delta \avg{J^\mu} = G^{\mu\nu} a_\nu^\mathsc{s} =  G^{\mu\nu} a_\nu^{(0)}\text{ .}
    \end{align}
    Since the bulk theory is left unchanged by the deformation, the Maxwell-field fluctuations $a_\mu$ satisfies the same linearized equations of motion with the same near-boundary behavior, the only difference is that now we interpret the leading-order coefficient $a_\nu^{(0)}$ not as the source alone but according to the right-hand side of eq.\ \eqref{eq:neumman_bdy}. In terms of the source $a_\mu^\mathsc{s}$ in the presence of a double-trace deformation, expression \eqref{eq:linear_response} now becomes
    \begin{align}\label{eq:response_double_trace}
      \delta \avg{J^\mu} = G^{\mu\nu} \left(a^\mathsc{s}_\nu - \alpha^2 V_{\nu\sigma}\delta\avg{J^\sigma}\right) \text{ ,}
    \end{align}
    where for convenience we defined $V_{\mu\nu} = (\eta_{\mu\nu} - (1 - \xi)k_\mu k_\nu/k^2)/{k^2}$. Rearranging the equation in order to put it into the form $\delta\avg{J^\mu} =\chi^{\mu\nu} a^\mathsc{s}_\nu$ to extract the response function $\chi^{\mu\nu}$ we have that
    \begin{align}
      \left(\delta^\mu_\sigma + \alpha^2 G^{\mu\nu}V_{\nu\sigma}\right)\delta \avg{J^\sigma} = G^{\mu\nu} a^\mathsc{s}_\nu \text{ ,}
    \end{align}
    and we see that the response function assumes the RPA-like form
    \begin{align}\label{eq:response_function}
      \chi = \left(I + \alpha^2 G V\right)^{-1}G \text{ .}
    \end{align}
    Explicitly, introducing $\Pi$ by means of
    \begin{align}
      G(\omega, \bm k) = \left(\begin{array}{cc}
        \bm k^2 \Pi(\omega, \bm k)  & \omega \bm k \Pi(\omega, \bm k) \\
        \omega \bm k \Pi(\omega, \bm k)  & \omega^2 \Pi(\omega, \bm k) \\
      \end{array}\right) \text{ ,}
    \end{align}
    we obtain
    \begin{align}\label{eq:rpa}
      \begin{split}
      &\chi(\omega, \bm k) = \\
      &\left[\left(
      \begin{array}{cc}
        1  & 0 \\
        0  & 1 \\
      \end{array}\right) + \frac{\alpha^2}{- \omega^2 + \bm k^2}  \left(\begin{array}{cc}
        \bm k^2 \Pi  & \omega \bm k \Pi \\
        \omega \bm k \Pi  & \omega^2 \Pi \\
      \end{array}\right) 
      \left(\begin{array}{cc}
        -1 - (1 - \xi) \frac{\omega^2}{-\omega^2 + \bm k^2}  & (1 - \xi) \frac{\omega \bm k}{-\omega^2 + \bm k^2} \\
        (1 - \xi) \frac{\omega \bm k}{-\omega^2 + \bm k^2} & 1 - (1 - \xi) \frac{\bm k^2}{-\omega^2 + \bm k^2} \\
      \end{array}\right)\right]^{-1}\\
      \\
      & \cdot \left(\begin{array}{cc}
        \bm k^2 \Pi  & \omega \bm k \Pi \\
        \omega \bm k \Pi  & \omega^2 \Pi \\
      \end{array}\right) = 
      \frac{1}{1 - \alpha^2 \Pi(\omega, \bm k)}  
      \left(\begin{array}{cc}
        \bm k^2 \Pi(\omega, \bm k)  & \omega \bm k \Pi(\omega, \bm k) \\
        \omega \bm k \Pi(\omega, \bm k)  & \omega^2 \Pi(\omega, \bm k) \\
      \end{array}\right)  \text{ ,}
    \end{split} 
  \end{align}
  that is of course independent of the gauge-choice parameter $\xi$. In particular, we can see that the density-density response function is
  \begin{align}\label{eq:density_rpa}
    \chi^{00}(\omega, \bm k) =  \frac{\bm k^2 \Pi(\omega, \bm k)}{1 - \alpha^2\Pi(\omega, \bm k)} = \frac{G^{00}(\omega, \bm k)}{1 - \frac{\alpha^2}{\bm k^2}G^{00}(\omega, \bm k) } \text{ ,}
  \end{align}
  as expected from RPA calculations.

  From relativistic hydrodynamics we know that in the low-energy limit $G^{00}$ assumes the form (\cite{Kovtun2012})
  \begin{align}\label{eq:hydro_density_density}
    G^{00}(\omega, \bm k) \simeq \frac{\avg{\rho}^2}{\avg{\epsilon} + \avg{P}} \frac{\bm k^2}{\omega^2 - \bm k^2 v_s^2} \quad \text{ for } \quad \omega \gtrsim \bm k \text{ ,}
  \end{align}
  with the equilibrium quantities defined in eqs.\ \eqref{eq:equilibrium_pressure_density}-\eqref{eq:equilibrium_density}, and $v_s^2$ the speed of sound squared $v_s^2 = \partial P/\partial{\epsilon} = 1/(d - 1)$. This suggest that the RPA-like response function in eq.\ \eqref{eq:density_rpa} contains a gapped plasmon mode, with the plasma frequency determined by 
  \begin{align}\label{eq:hydro_plasma_frequency}
    \omega_p^2 = \alpha^2 \frac{(d - 1)\avg{\rho}^2}{d \avg{\epsilon}} = \alpha^2 \lim_{\omega \to 0} G^{xx}(\omega, \bm k = \bm 0)\text{ .}
  \end{align}

\begin{figure}[h]\centering
  \begin{subfigure}{.5\textwidth}
    \centering
    \includegraphics[width=\linewidth]{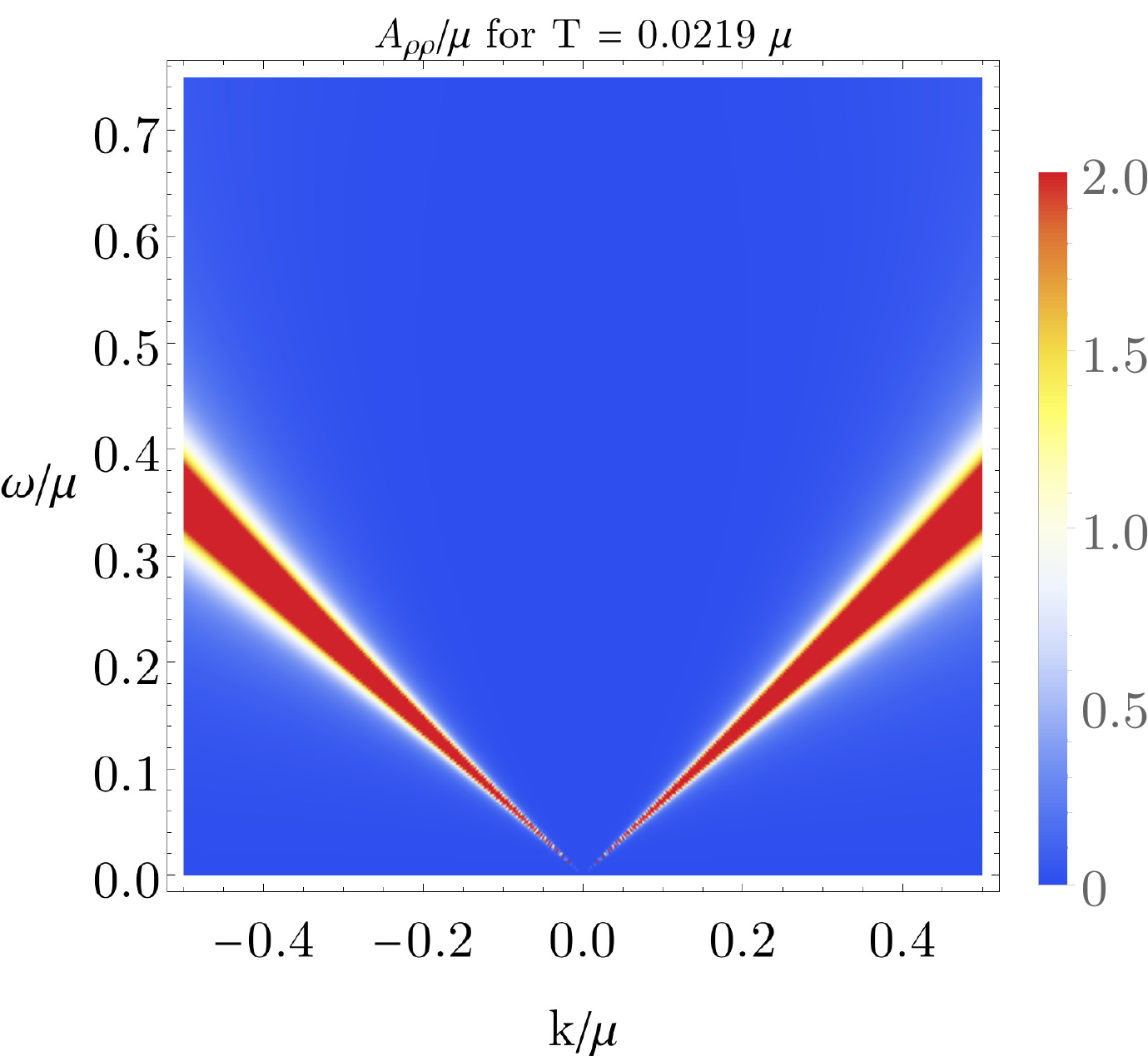}
  \end{subfigure}%
\begin{subfigure}{.5\textwidth}
  \centering
  \includegraphics[width=\linewidth]{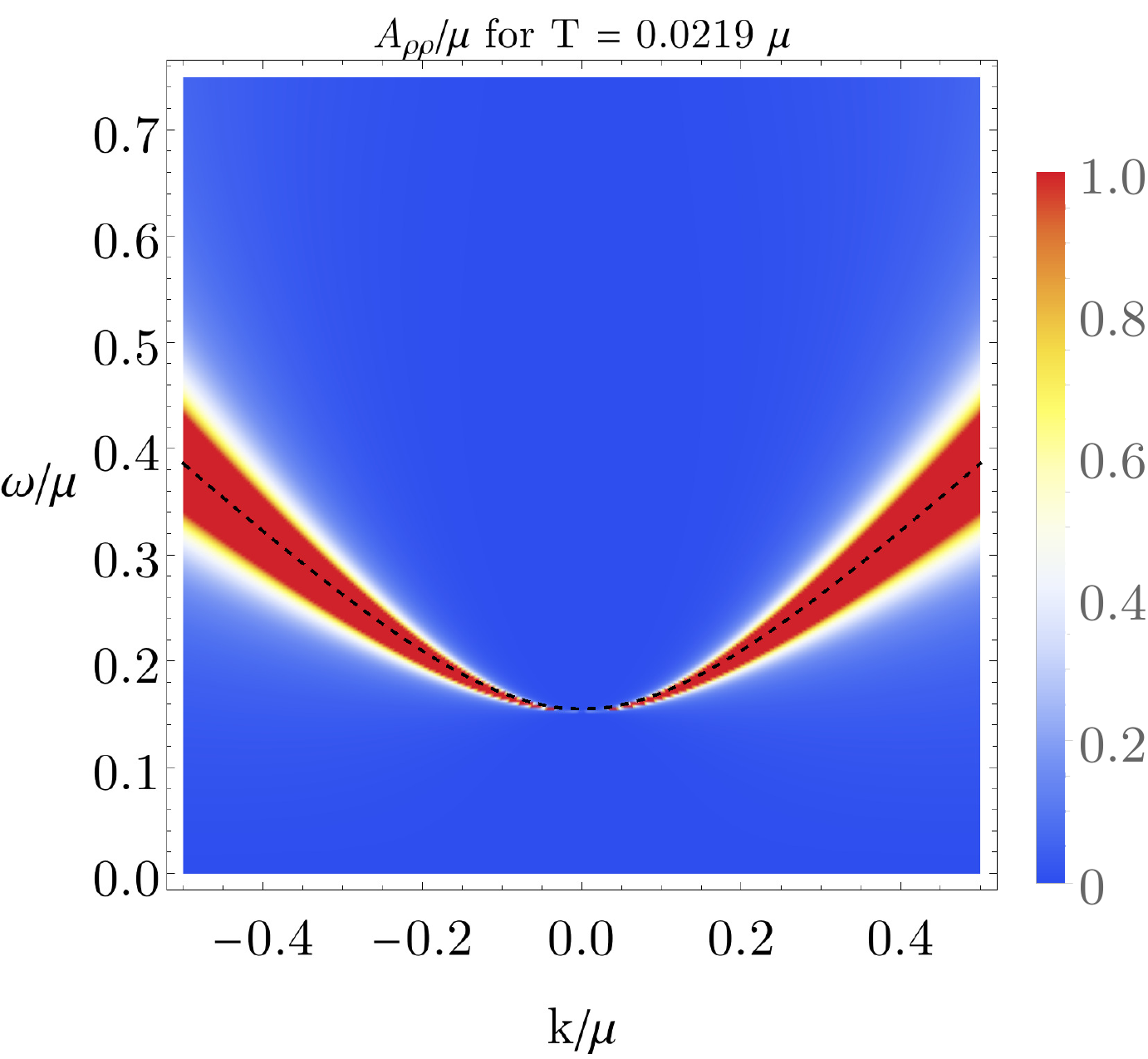}
\end{subfigure}
\vskip\baselineskip
 \begin{subfigure}{.5\textwidth}
    \centering
    \includegraphics[width=\linewidth]{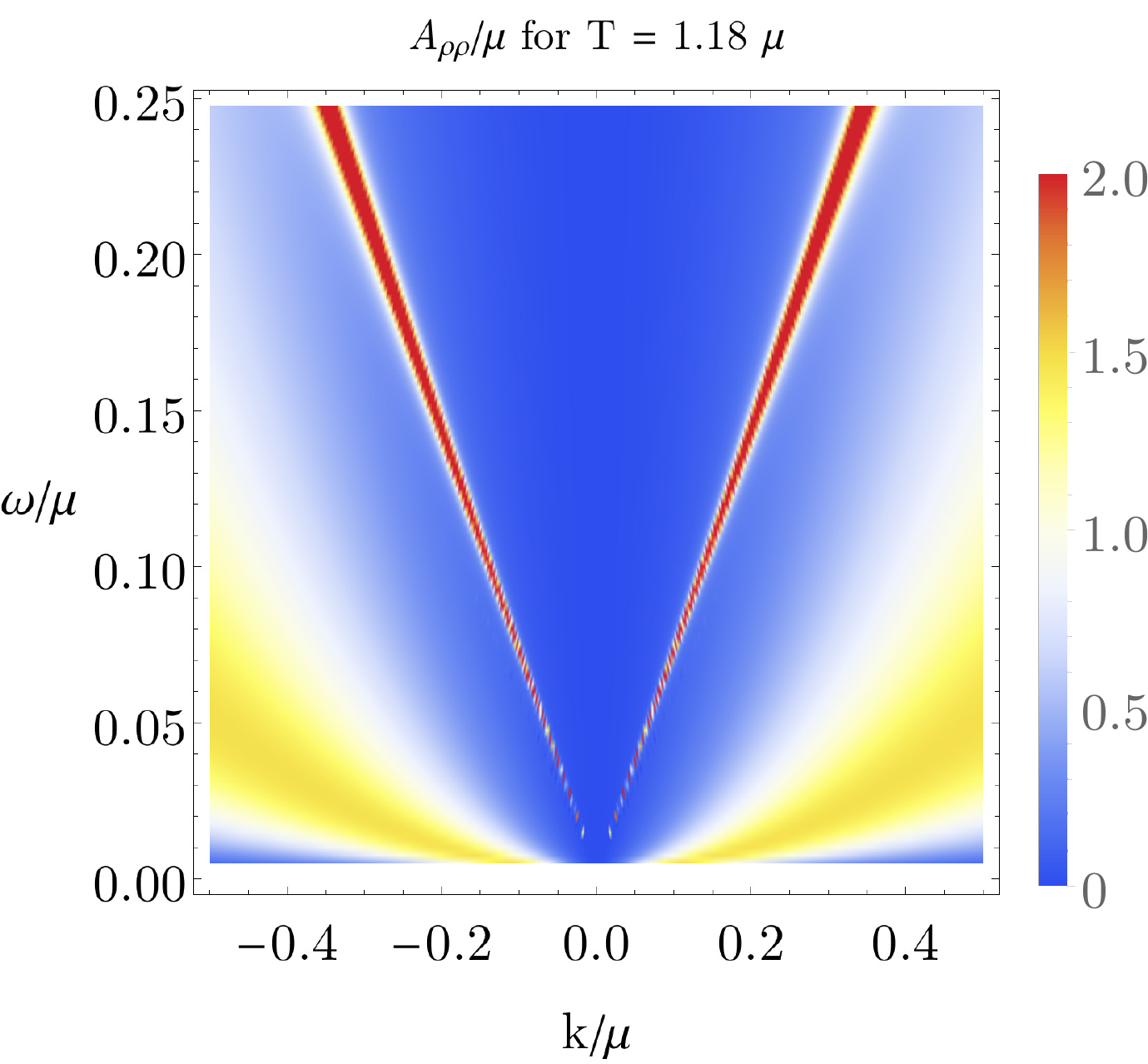}
  \end{subfigure}%
\begin{subfigure}{.5\textwidth}
  \centering
  \includegraphics[width=\linewidth]{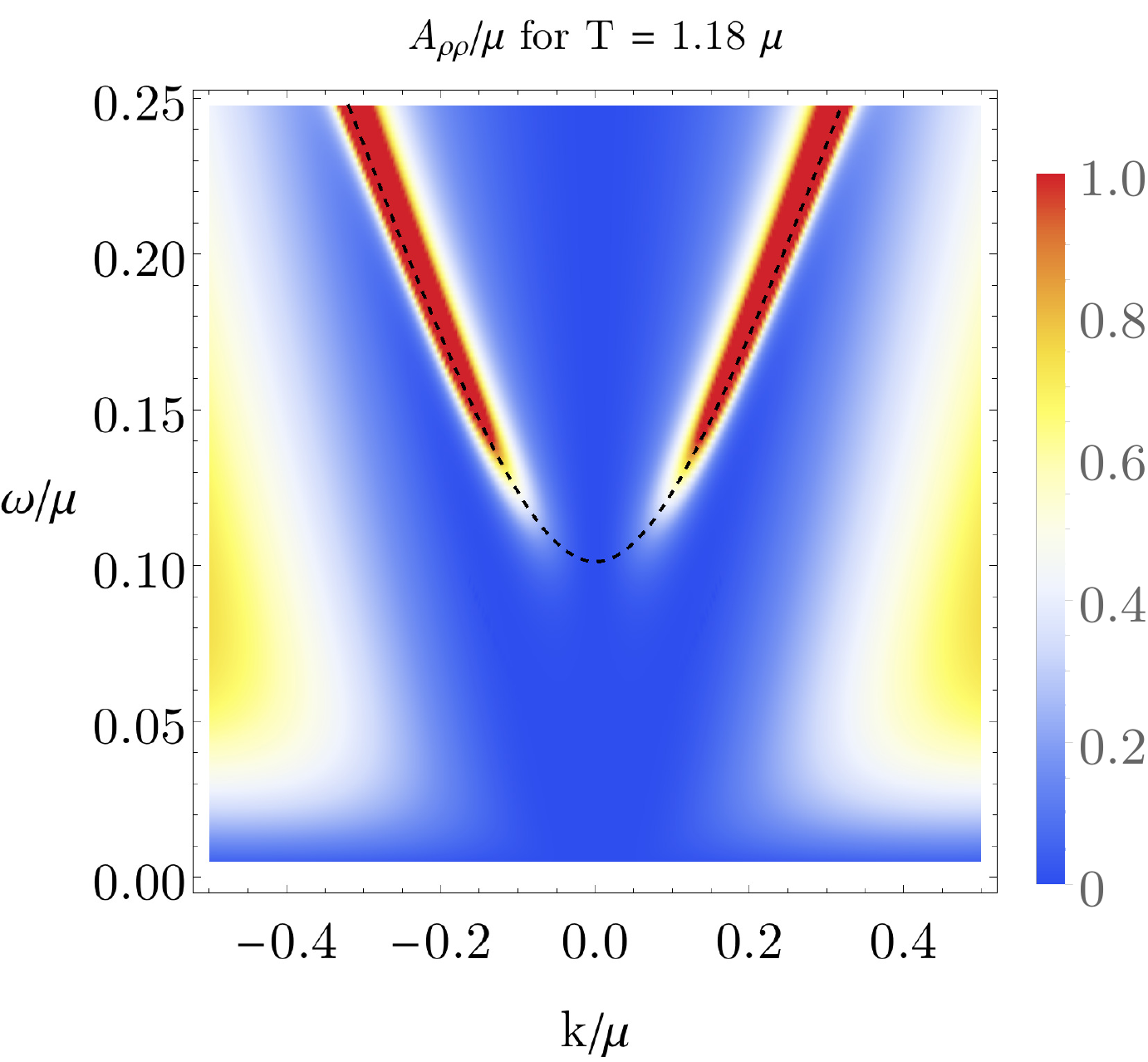}
\end{subfigure}
\caption{\label{fig:gapped_plasmon} The density spectral function for $T \simeq 0.02 \mu$ (top) and $T \simeq 1.18 \mu$ (bottom) without (left) and with (right) double-trace deformation. Here we have taken $\alpha^2 = 1/25$. The black dashed line represents the plasmon modes computed in the hydrodynamic approximation. We see that the double-trace deformation turns the low-energy sound modes into a gapped plasmon mode, although the low-energy excitation are less visible as the height of the peak for low $\abs{\bm k}/\mu$ becomes smaller for higher values of $T/\mu$.}
\end{figure}

\begin{figure}[h]\centering
  \begin{subfigure}{.5\textwidth}
    \centering
    \includegraphics[width=\linewidth]{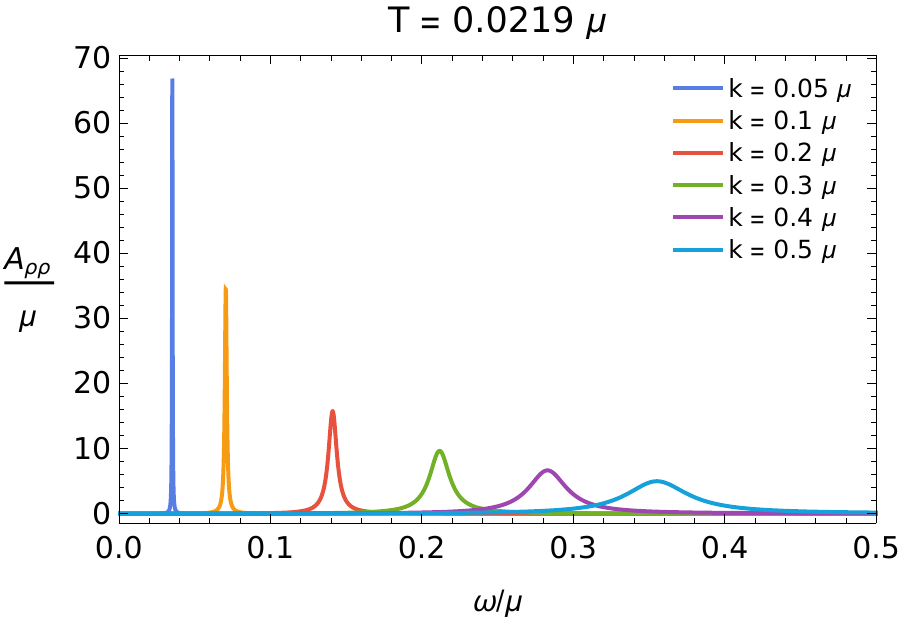}
  \end{subfigure}%
\begin{subfigure}{.5\textwidth}
  \centering
  \includegraphics[width=\linewidth]{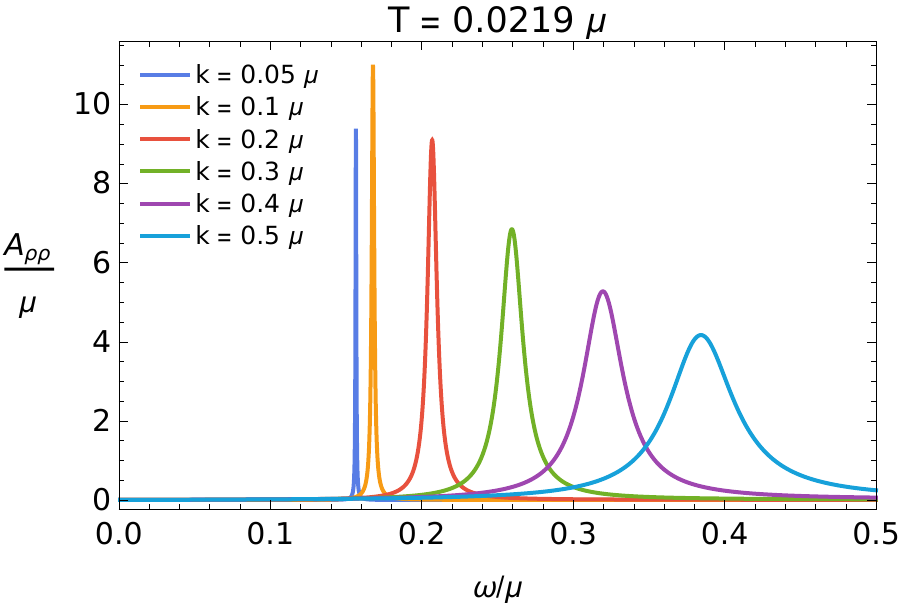}
\end{subfigure}
\vskip\baselineskip
 \begin{subfigure}{.5\textwidth}
    \centering
    \includegraphics[width=\linewidth]{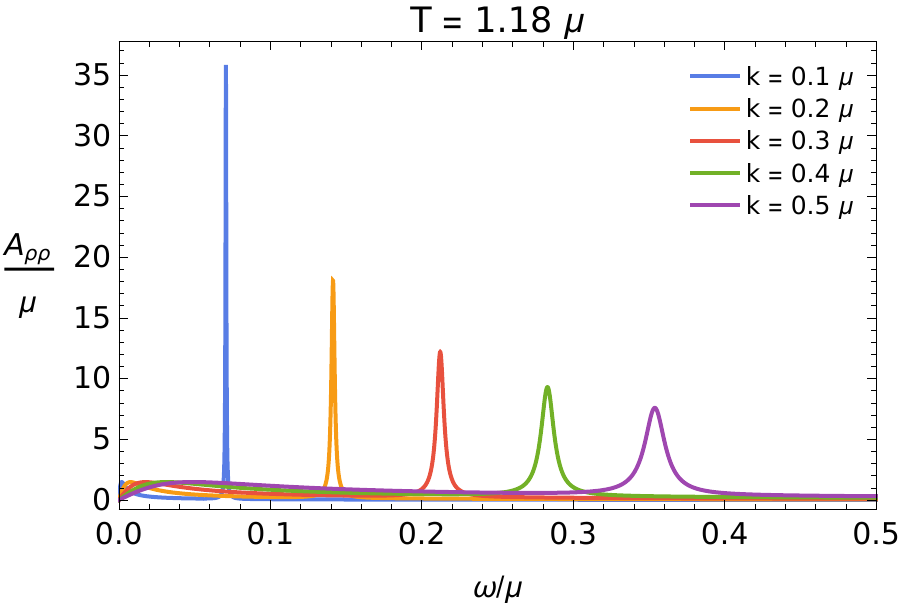}
  \end{subfigure}%
\begin{subfigure}{.5\textwidth}
  \centering
  \includegraphics[width=\linewidth]{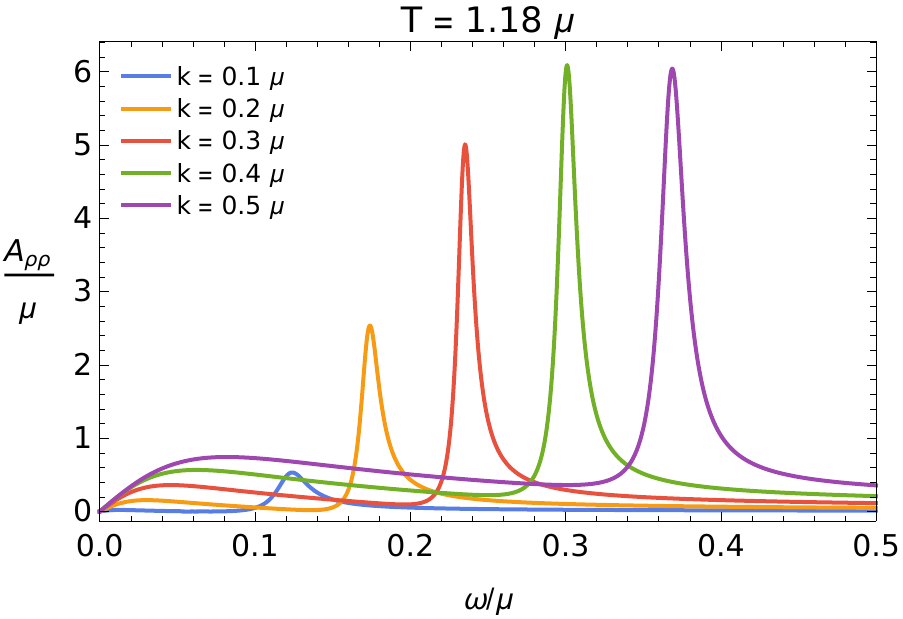}
\end{subfigure}
\caption{\label{fig:slice_gapped_plasmon} The density spectral function for $T \simeq 0.02 \mu$ (top) and $T \simeq 1.18 \mu$ (bottom) without (left) and with (right) double-trace deformation for different values of $k/\mu$ ($\alpha^2 = 1/25$). We see that the height of the peak of the plasmon modes is lower than in the corresponding sound modes due to screening of the charged particles. Moreover we observe that for low momenta, the peak in the RPA response function decreases as we move towards smaller values of $\abs{\bm k}/\mu$.}
\end{figure}

\begin{figure}
\centering 
\includegraphics[width=.49\textwidth]{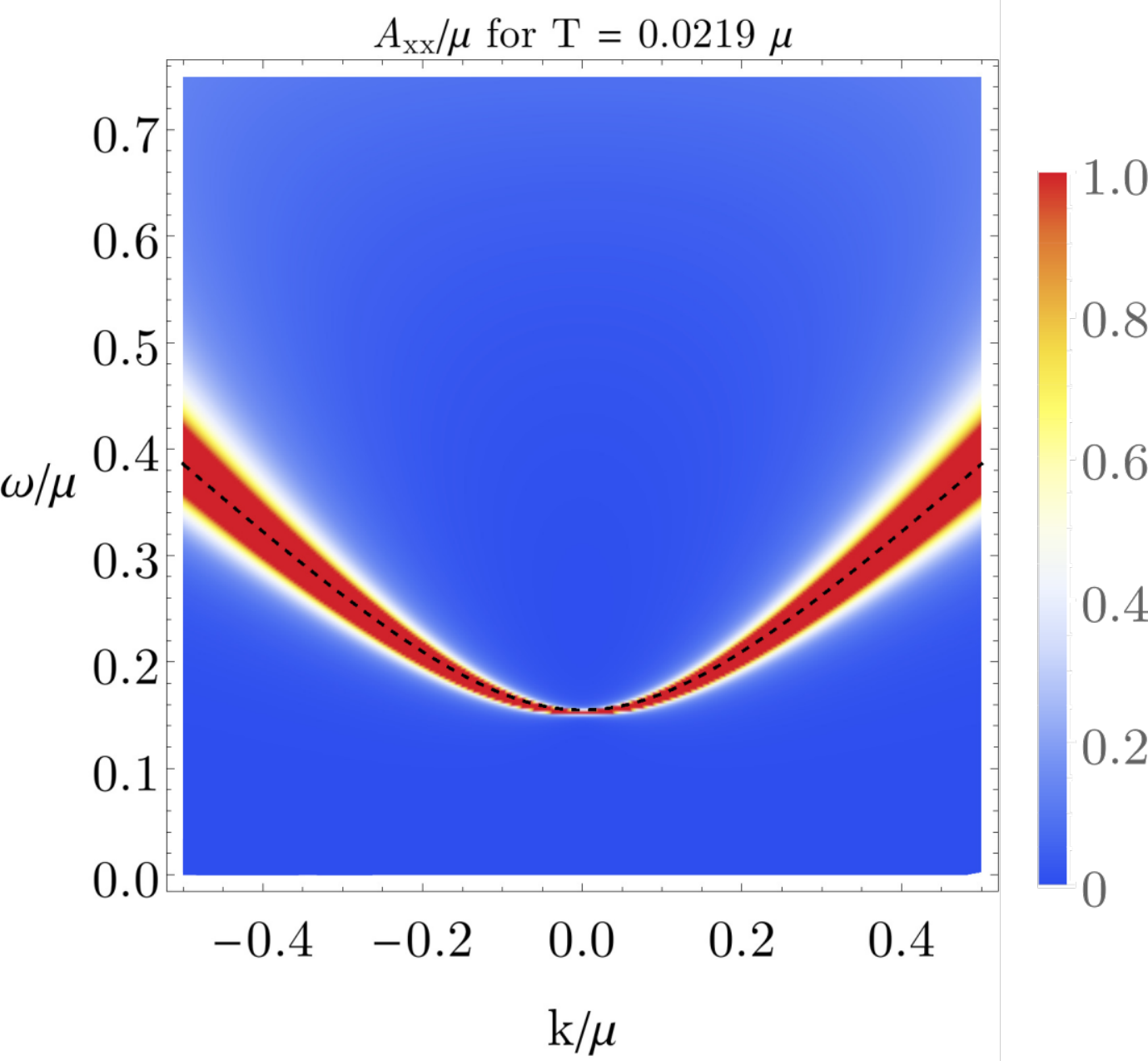}
\hfill
\includegraphics[width=.49\textwidth]{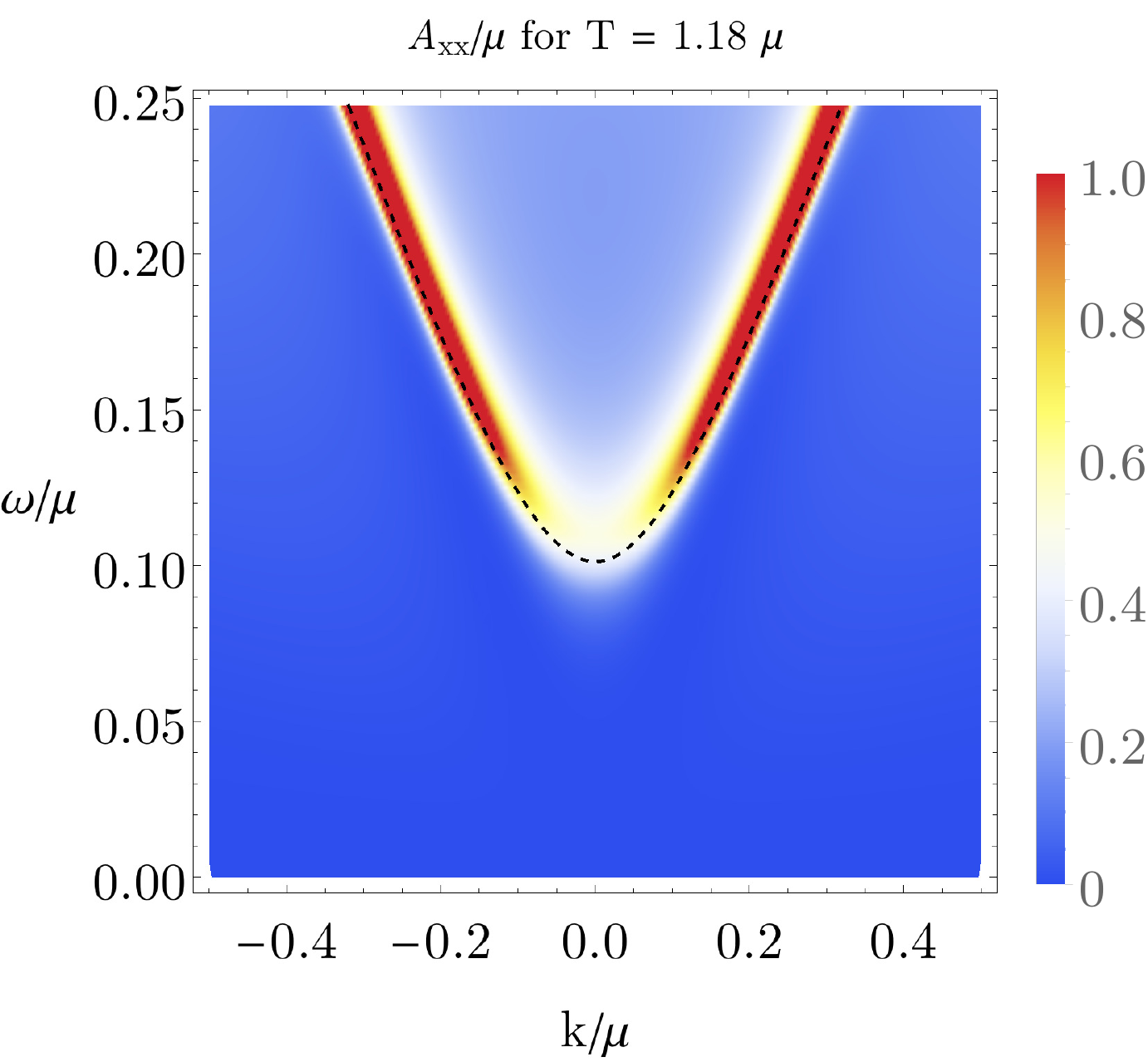}
\caption{\label{fig:gapped_plasmon_xx} The current spectral function for $T \simeq 0.02 \mu$ (left) and $T \simeq 1.18 \mu$ (right) with a double-trace deformation ($\alpha^2 = 1/25$). Here we can clearly see the gapped plasmon mode, as the height of the peak does not go to zero as $\bm k \to \bm 0$.}
\end{figure}

  In figure \ref{fig:gapped_plasmon}, we show side by side the density-density spectral function with and without double-trace deformation, for two different values of the parameter $T/\mu$, where we see that the linear sound modes turn into a gapped plasmon mode. The dashed black line represent the hydrodynamic approximation of the plasmon mode from eq.\ \eqref{eq:hydro_density_density}. We can think of higher values of $T/\mu$ as moving towards the zero chemical potential limit, where there is no plasmon mode. By comparing the two figures we can indeed observe that for higher values of $T/\mu$ the low-energy gapped mode becomes less visible in the spectral function. This can be seen in more details by comparing with figure \ref{fig:slice_gapped_plasmon}, where we show some slices of the spectral function for different fixed values of momentum $\abs{\bm k}/\mu$. Here we also notice that the height of the peak of the plasmon modes is lower than the corresponding peak in the Reissner-Nordstr\"om solution. This is due to the screening effect of the charged particles that opposes density fluctuations. Moreover, we see that the plasmon peak initially increases as we move to smaller values of $\abs{\bm k}/\mu$, as was the case for the linear modes, but it then starts to decrease for small value of $\abs{\bm k}/\mu$, due to the $\bm k^2$ factor in the low-energy limit of the density-density spectral function. Even though in this paper we mainly focus on the density-density response, the plasmon modes are of course also present in the current-current spectral function as we show in figure \ref{fig:gapped_plasmon_xx}. Here the plasmon mode is more easily visible as there is no $\bm k^2$ suppression, since from Ward's identities we know that $\bm k^2 G^{xx} = \omega^2 G^{00}$. 
  
This gapped plasmon mode is observed in $(3+1)$-dimensional metals, however, in an experimental $(2+1)$-dimensional system we expect to find plasmon modes of the form $\omega \propto \sqrt{\abs{\bm k}}$. Notice that the expression for the RPA-like correction to the response function in eq.\ \eqref{eq:density_rpa} has been derived in an arbitrary spacetime dimension and it is independent of the value of $d$. This implies that in any dimension, the qualitative behavior of the spectral function looks the same as the one in figure \ref{fig:gapped_plasmon}, although the position and residue of the poles of course change. In the next section we show how to modify the double-trace deformation to, from an experimental point of view, correctly describe plasmon modes in $(2 + 1)$-dimensional systems.

\section{Plasmon modes in $d = 2 + 1$}\label{sec:2dplasmon}
  
  As already noted in ref.\ \cite{Aronsson2018}, the presence of a gapped mode is due to the fact that in the holographic theory of a $(2+1)$-dimensional model, we also constrain the boundary dynamical photons to live in $2 + 1$ dimensions, while in layered materials studied in laboratories, the current is constrained on the $(2+1)$-dimensional layer, but the photons are free to move in all the spacetime dimensions, giving rise to an effective Coulomb potential $V(\bm k)\propto 1/\abs*{\bm k}$. Put differently, the Coulomb potential between charges at positions $\bm x$ and $\bm x'$ behaves as $1/\abs{\bm x - \bm x'}$  and not as $\log\abs{\bm x - \bm x'}$.

  \subsection{Single Layer}
  In order to model a more realistic $(2+1)$-dimensional system, where photons that mediate the Coulomb interaction are free to move in the whole three-dimensional space, we consider a $d  = 2+1$ boundary theory, dual to the $(d+1)$-dimensional bulk theory from eq.\ \eqref{eq:dimless_einstein-maxwell_action}, but we define the deformation of the boundary theory by starting from a $(d+1)$-dimensional boundary term. Defining the $(d+1)$-dimensional vector $(x, z)$ and momentum $(k, k_z)$, with $x$ and $k$ living in $d$ spacetime dimensions while $z$ and $k_z$ represent the extra spatial dimension normal to the layer and the associated momentum respectively, we restrict the current operator to a plane as
  \begin{align}\label{eq:current_on_plane}
    \begin{cases}
      J^\mu(x, z) = J^\mu(x) \delta(z), & \mu = 0, \dots, d - 1\\
      J^z(x, z) = 0 &
    \end{cases}\text{ .}
  \end{align} 

  Inserting this expression into eq.\ \eqref{eq:maxwell_boundary}, Fourier transforming and performing the Gaussian integral over the Maxwell field we obtain
  \begin{align}\label{eq:RPA_potential_2d}
    S_{\text{M}} = -\frac{1}{2}\int\frac{\dif^{d} k}{(2\pi)^d} \int \frac{\dif k_z }{2\pi}\, \alpha^2 \left(\avg{J^\mu(k)} \frac{\eta_{\mu\nu}}{k^2 + k_z ^2} \avg{J^\nu(-k)} \right) \text{ ,}
  \end{align}  
  that can be integrated over $k_z$ to obtain the $d$-dimensional boundary deformation
  \begin{align}
    -\frac{1}{2}\int \frac{\dif^{d} k}{(2 \pi)^d} \, \frac{\alpha^2}{2} \left(\avg{J^\mu(\omega, \bm k)} \frac{\eta_{\mu\nu}}{\sqrt{-\omega^2 + \bm k^2} } \avg{J^\nu(-\omega, -\bm k)} \right) \quad \text{ for } \bm k^2 > \omega^2 \text{ .}
  \end{align}
  This in particular implies that the response function now takes the form:
  \begin{align}\label{eq:RPA_2d}
    \chi(\omega,\bm k) = \frac{1}{1 - \frac{\alpha^2}{2} \sqrt{-\omega^2 +\bm k^2} \Pi(\omega, \bm k)}  
    \left(\begin{array}{cc}
        \bm k^2 \Pi(\omega, \bm k)  & \omega\bm k \Pi(\omega, \bm k) \\
        \omega \bm k \Pi(\omega, \bm k)  & \omega^2 \Pi(\omega, \bm k) \\
      \end{array}\right)  \text{ .}
  \end{align}
  However, we are ultimately interested in describing condensed-matter systems, where the Fermi velocity is considerably smaller than the speed of light, so $v_F \ll c$ and we look into a regime where $\abs{\bm k} \gg \omega$ in eq.\ \eqref{eq:RPA_2d}. As a result, the density-density response function is well approximated by the familiar form 
  \begin{align}
    \chi^{00}(\omega, \bm k) = \frac{G^{00}(\omega,\bm k)}{1 - \frac{\alpha^2}{2\abs*{\bm k}} G^{00}(\omega, \bm k)} \text{ ,}
  \end{align}
  that is expressed in terms of the static Coulomb potential expected in two-dimensional metals. Notice that the coupling constant for the single layer is $\alpha^2/2$, in accordance with eq.\ \eqref{eq:rpa_cond_matt}.

\begin{figure}
  \centering 
  \includegraphics[width=.49\textwidth]{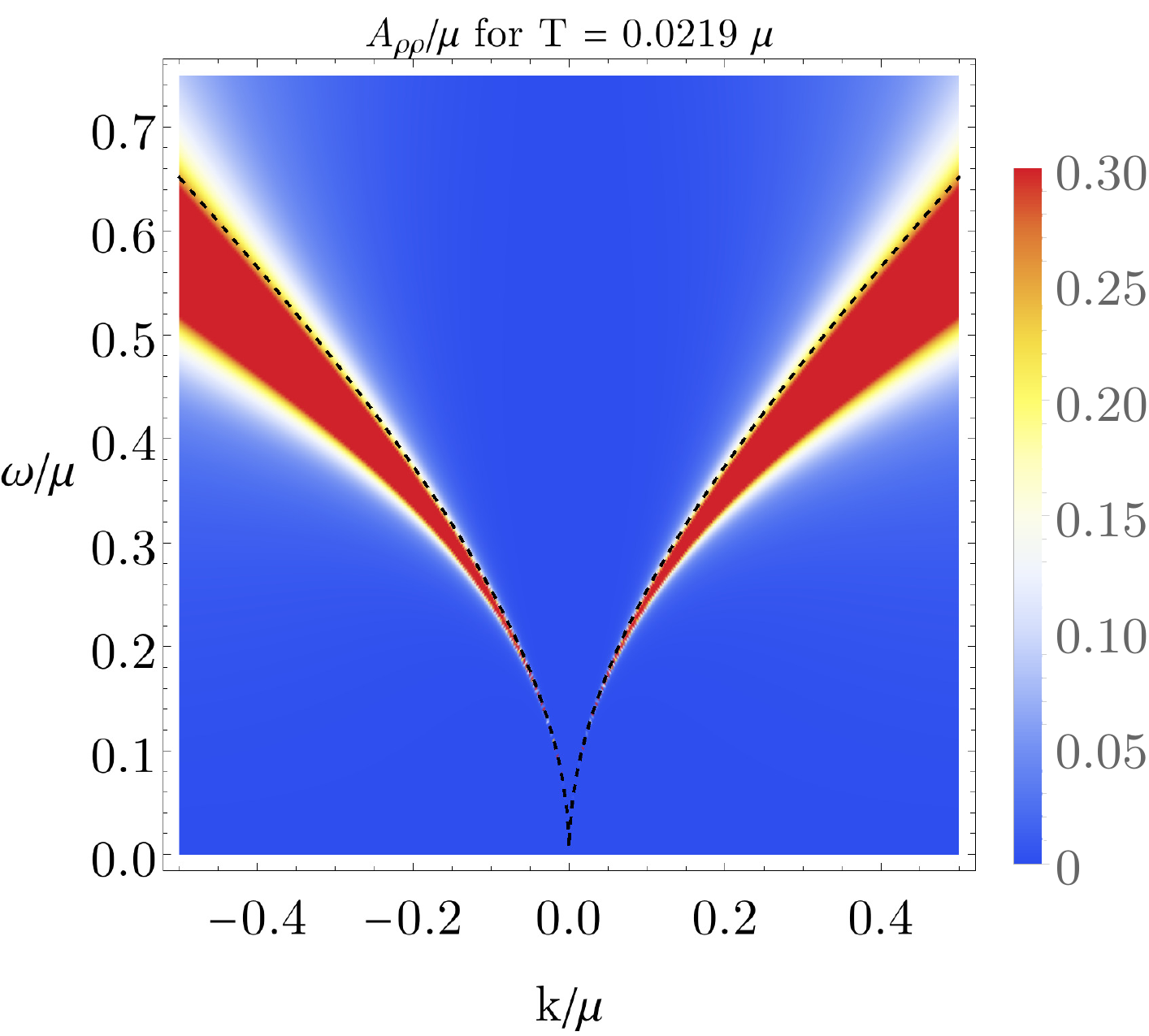}
  \hfill
  \includegraphics[width=.49\textwidth]{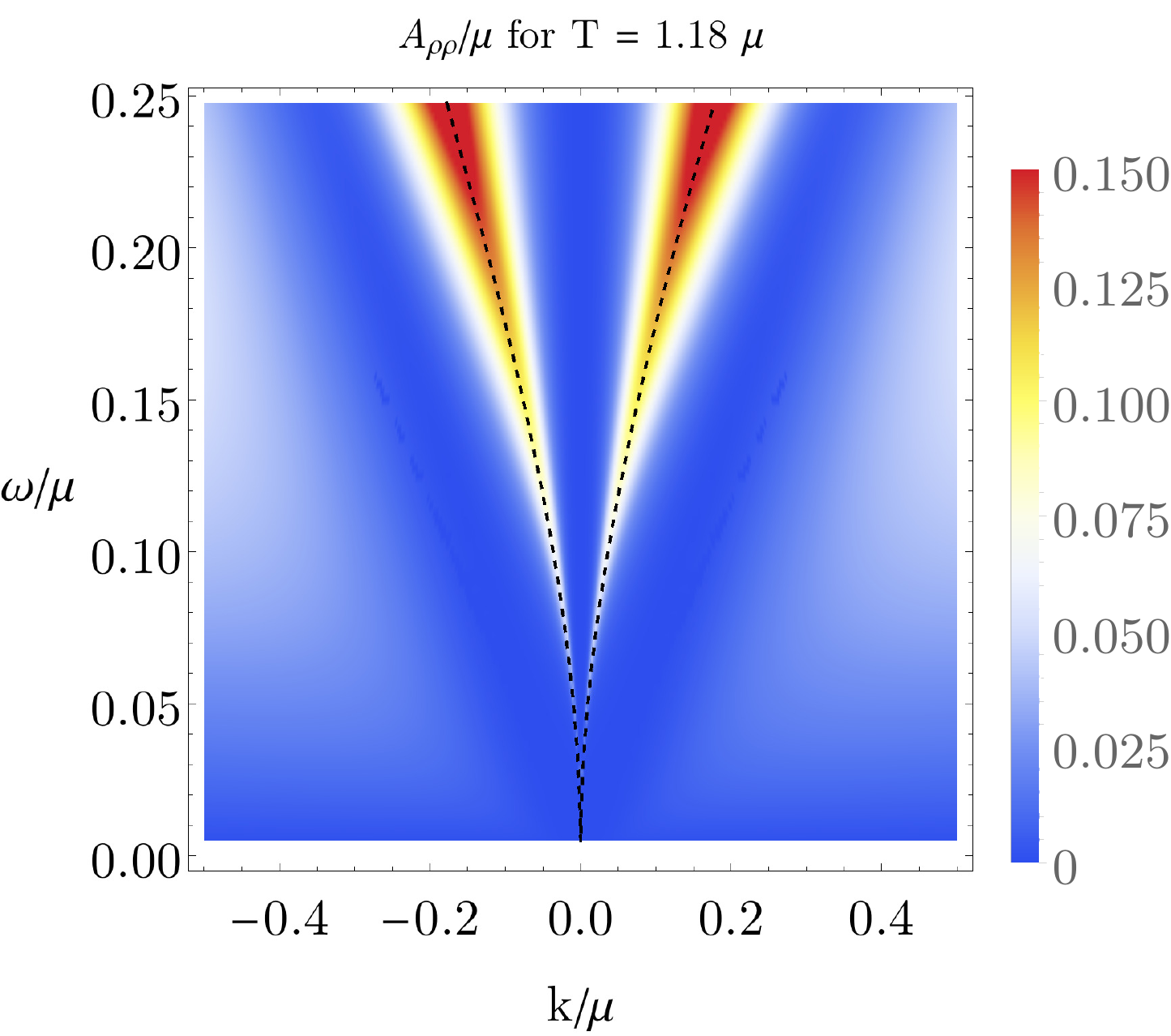}
  \caption{\label{fig:sqrt_plasmon_full} The density spectral function for $T \simeq 0.02 \mu$ (left) and $T \simeq 1.2 \mu$ (right). We have taken $\alpha^2/2 = 1$. Here we see that the sound modes from the Reissner-Nordstr\"om solution turn into $\omega \propto \sqrt{\abs{\bm k}}$ modes at low energies. The dashed black line represent the hydrodynamic solution for the dispersion relation $\omega^2 = 2\avg{\rho}^2 \abs{\bm k} /3\avg{\epsilon}$}
  \end{figure}

\begin{figure}
  \centering 
  \includegraphics[width=.49\textwidth]{images/fixedK_rn_T0_022.pdf}
  \hfill
  \includegraphics[width=.49\textwidth]{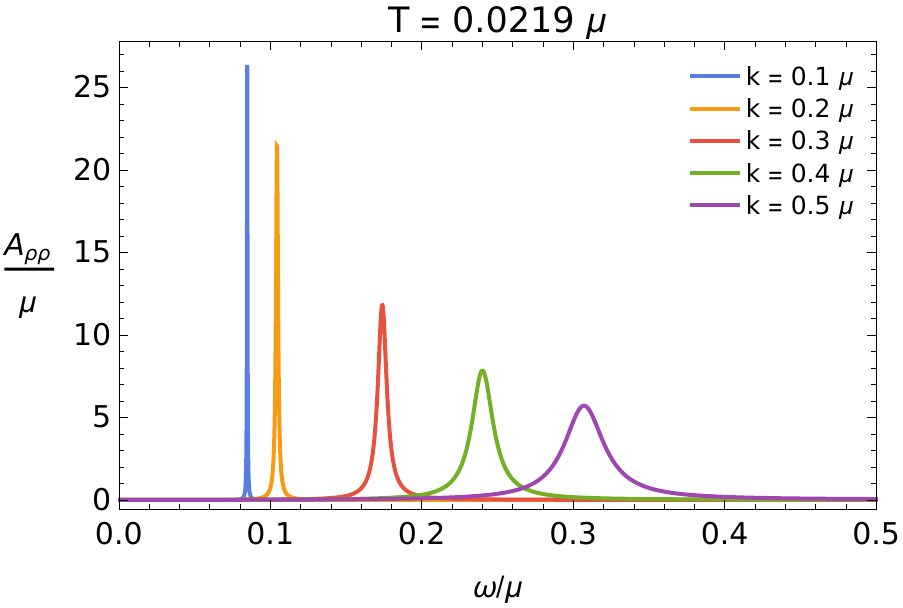}
  \caption{\label{fig:slice_sqrt_plasmon_002} The density spectral function for different values of $\abs{\bm k}/\mu$  in the Reissner-Nordstr\"om solution (left) and with Coulomb corrections (right). We see that due to screening in the charged system the plasmon peaks shift towards higher frequencies and are lower in height than the corresponding peak of the sound mode.}
  \end{figure}

\begin{figure}[h]\centering
    \includegraphics[width=0.65\textwidth]{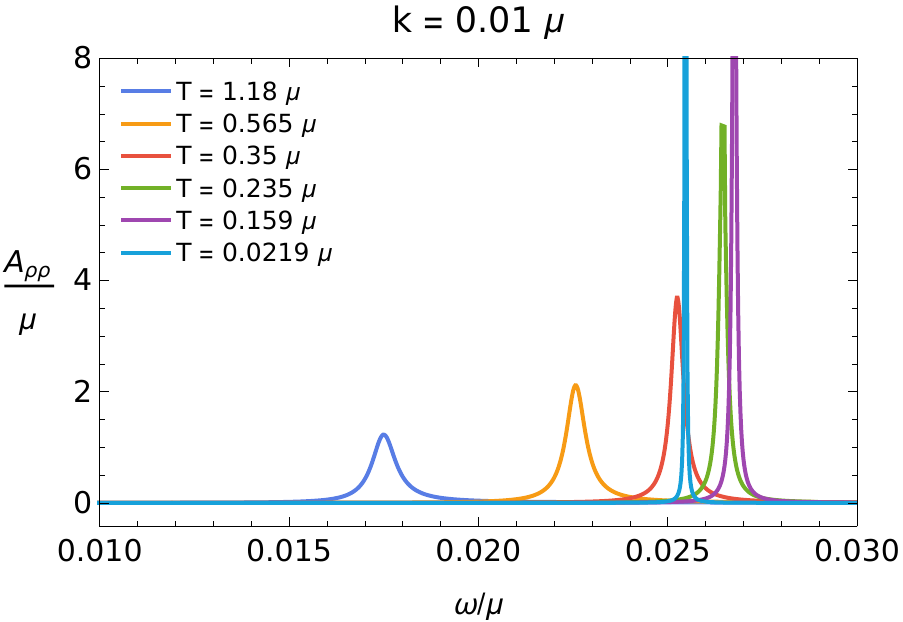}
     \caption{Temperature dependence of the plasmon mode for $\abs{\bm k}/\mu = 0.01$ and $\alpha^2/2 = 1/10$. We see that the peaks become less pronounced as we raise the temperature due to the effect of temperature fluctuations. Moreover, for high temperatures, the position of the plasmon mode shifts towards the position of the sound mode in the neutral system, as temperature fluctuations start dominating over the Coulomb interaction. As we lower the temperature we see that the plasma frequency reaches a maximum at $T/\mu = 1/2\pi$ (purple), before moving to lower values as the temperature is decreased further.}
     \label{fig:sqrt_plasmon_Tdep}
  \end{figure}

\begin{figure}[h]\centering
    \includegraphics[width=0.65\textwidth]{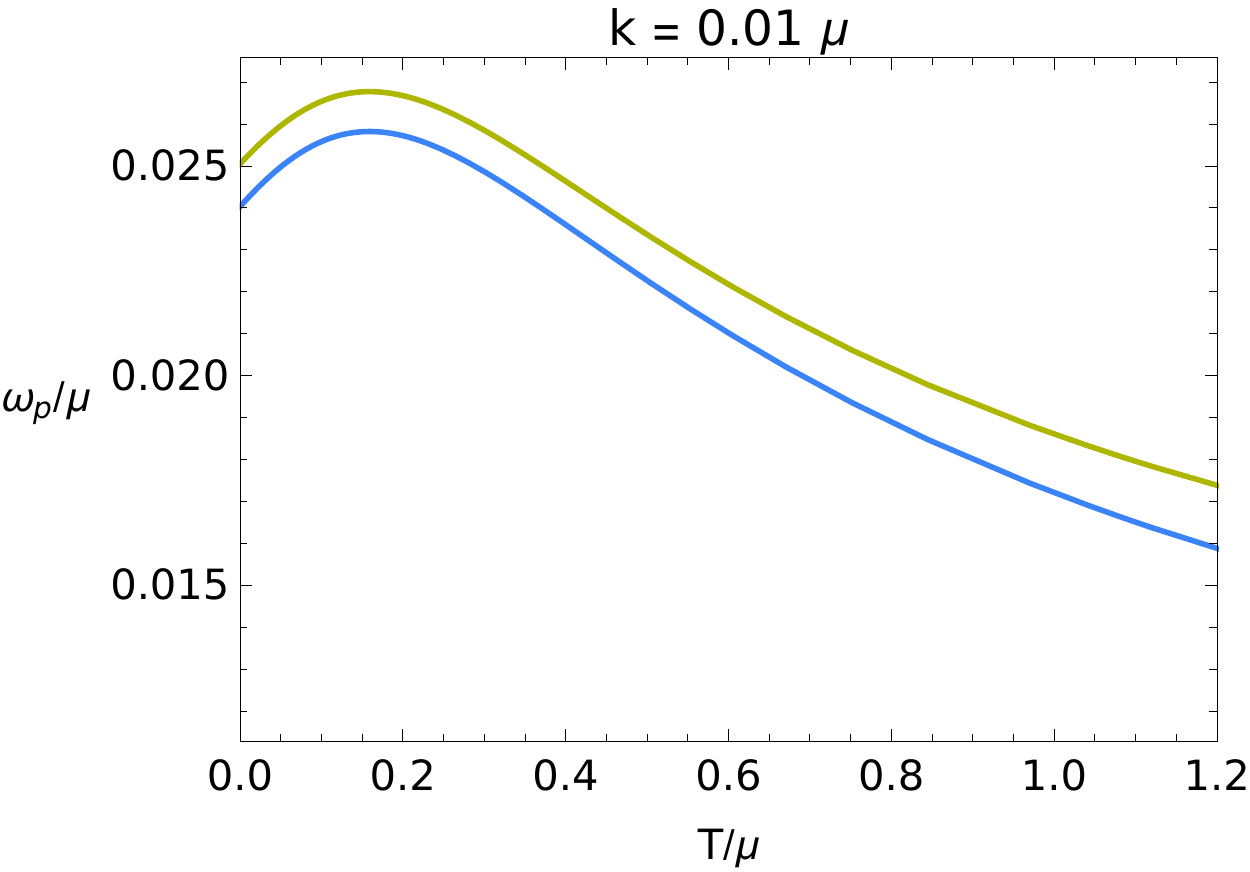}
     \caption{Temperature dependence of the hydrodynamic plasma frequency for $\abs{\bm k}/\mu = 0.01$ and $\alpha^2/2 = 1/10$, with (green) and without (blue) a higher-order momentum correction $v_s^2 \bm k^2$. We see that the plasma frequency has a maximum at $T/\mu = 1/2\pi$, corresponding to $\omega_p \simeq 0.0268 \mu$.}
     \label{fig:hydro_plasma_frequency}
  \end{figure}

  In figure \ref{fig:sqrt_plasmon_full} we see, for two different values of $T/\mu$, how this form of the potential does indeed give rise to $\omega \propto \sqrt{\abs{\bm k}}$ modes observed in a (spatially) two-dimensional system such as graphene \cite{Lucas2018}. As before, the black dashed lines show the hydrodynamic plasmon modes 
  \begin{align}\label{eq:plasma_frequency_sqrt}
    \omega_p^2 = \frac{\alpha^2}{2} \frac{2\avg{\rho}^2}{3 \avg{\epsilon}} \abs{\bm k} \text{ .}
  \end{align}
  In figure \ref{fig:slice_sqrt_plasmon_002}, we observe the momentum dependence of the peak for $T \simeq0.02 \mu$, where again we notice that the height of the peak in the charged system is less pronounced with respect to the one in the Reissner-Nordstr\"om solution, due to the screening effect of charged particles. 
  Furthermore, in figure \ref{fig:sqrt_plasmon_Tdep}, we study the temperature dependence of this mode for a fixed value of $\abs{\bm k}/\mu$. In the Reissner-Nordstr\"om solution, the position of the peak for low enough momenta is independent of the temperature and given by the speed of sound $\omega = v_s \abs{\bm k} = \abs{\bm k}/\sqrt{2}$. In contrast, when we introduce Coulomb interactions we can see that, while the peak still gets sharper and the height increases as we lower the temperature,  the position of the peak is temperature dependent. For high temperatures, as the temperature is raised the mode shifts towards the position of the peak of the neutral solution ($\omega/\mu \simeq 0.035$ for $\abs{\bm k}/\mu = 0.05$), since temperature fluctuations start to dominate over the effect of Coulomb interactions. As we decrease the temperature, however, while the peak initially shifts to higher frequencies, it reaches a maximum frequency for $T/\mu = 1/2\pi$ (purple line in the plots) before starting to move back on the frequency axis as we further lower the temperature. This behavior is in accordance with the hydrodynamic prediction for the plasma frequency in eq.\ \eqref{eq:plasma_frequency_sqrt}. In figure \ref{fig:hydro_plasma_frequency} we show the temperature dependence of this hydrodynamic prediction in the grand-canonical ensemble, that is given by the temperature dependence of the thermodynamic quantities in eqs.\ \eqref{eq:equilibrium_pressure_density}-\eqref{eq:equilibrium_density}.  The green line includes the higher-order correction expressed in terms of the temperature-independent sound velocity $v_s = 1/\sqrt{2}$ 
  \begin{align}\label{eq:plasma_frequency_sqrt_sound}
    \omega_p^2 = \frac{\alpha^2}{2} \frac{2\avg{\rho}^2}{3 \avg{\epsilon}} \abs{\bm k} + v_s^2 \bm k^2 \text{ .}
  \end{align}
  The hydrodynamic plasma frequency shows a maximum at $T/\mu = 1/2\pi$, given by 
  \begin{align}
    \omega_p = \sqrt{\frac{\alpha^2}{3} \abs*{\bm k} \mu +  \frac{\bm k^2}{2}} \text{ ,}
  \end{align}
  that, for $\alpha^2/2 = 1/10$ and $\abs*{\bm k} = \mu/100$, gives $\omega_p \simeq 0.0268 \mu$, in agreement with the holographic results shown in figure \ref{fig:sqrt_plasmon_Tdep}. On the other hand, as $T/\mu \to \infty$, $\avg{\rho}^2/\avg{\epsilon} \to 0$ and we recover the sound dispersion.

  \subsection{Layered system}\label{subsec:layered}

    In a similar fashion to what we did in the last section, we present here a toy model for an infinite stack of $(2+1)$-dimensional layers. We show that we recover the form of the Coulomb potential for a layered electron gas \cite{Turlakov2003,Fetter1974}, and we present the density spectral function for a double-trace deformation of this form. The latter shows a dispersion relation for the low-energy plasmon excitations that changes from linear to a gapped mode as a function of Bloch momentum $p$ in the direction perpendicular to the layers. This behavior of the plasmon modes has very recently been observed in high-temperature cuprate superconductors consisting of stacked conducting copper-oxide layers \cite{Hepting2018}.

    For our toy model, we make the approximation that there are both strong in-plane short-range interactions and a long-range Coulomb interaction that couples the different layers.
    We, therefore,  model each single layer as an independent $d = 2 + 1$ boundary system dual to the $3 + 1$ Reissner-Nordstr\"om bulk theory, but we define a $(3+1)$-dimensional current as
    \begin{align}\label{eq:current_stack}
      \begin{cases}
        J^\mu( x, z) = \sum_n J^\mu( x, z) \delta(z - n \ell), & \mu = 0, \dots, d - 1\\
        J^z( x, z) = 0 & 
      \end{cases} \text{ ,}
    \end{align} 
    with $n \in \mathbb{Z}$ the layer index. The layers are stacked along the $z$ axis with $\ell$ the distance between each layer, and $J^\mu( x, n \ell)$ is the boundary operator computed from holographic calculations with sources $a^{(0)}_n$ not necessarily equal to each other. 

\begin{figure}[h]\centering
      \begin{subfigure}{.5\textwidth}
        \centering
        \includegraphics[width=\linewidth]{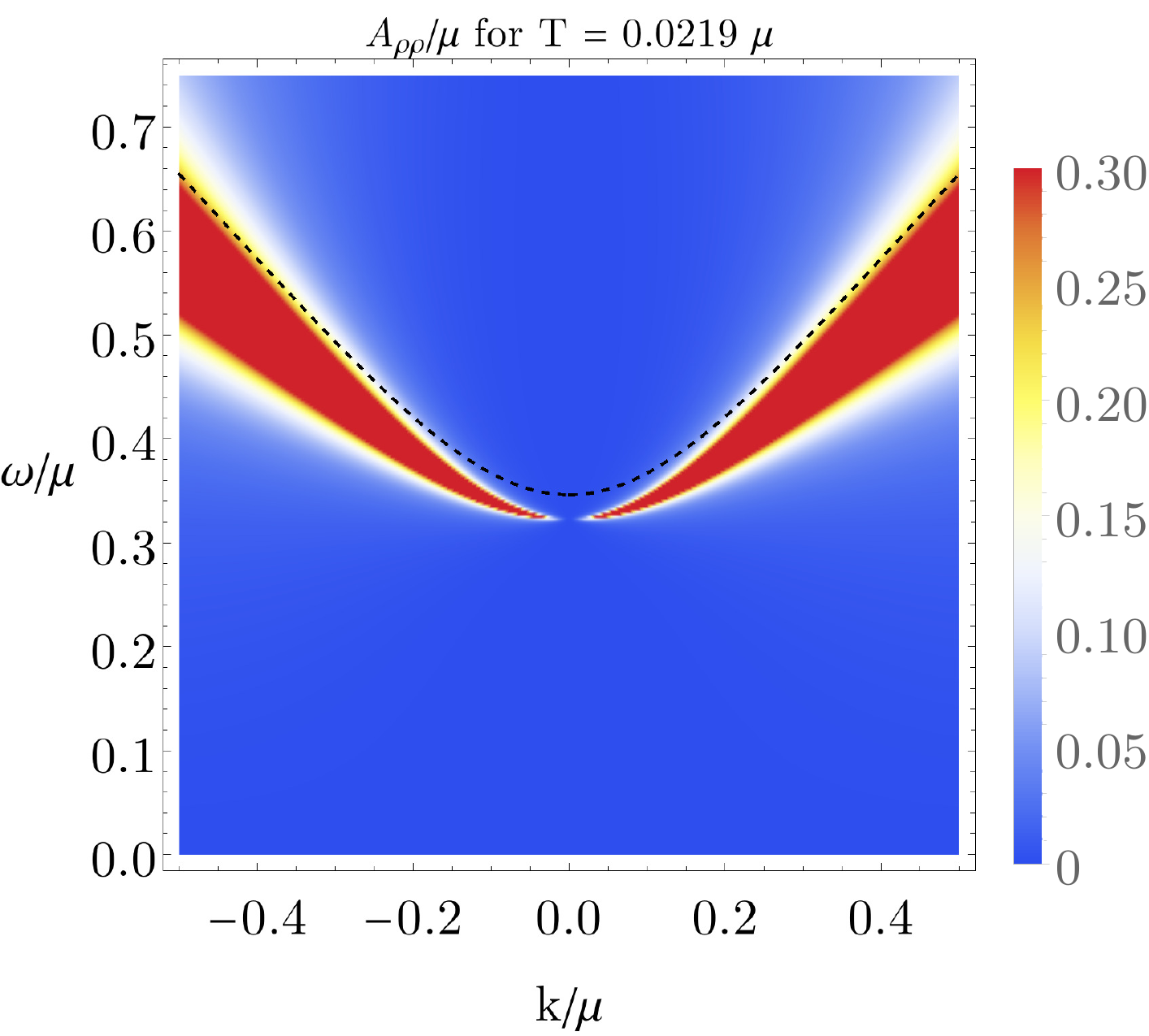}
        \caption{$p\ell = 0$}
        \label{fig:layered_cos_2pi}
      \end{subfigure}%
    \begin{subfigure}{.5\textwidth}
      \centering
      \includegraphics[width=\linewidth]{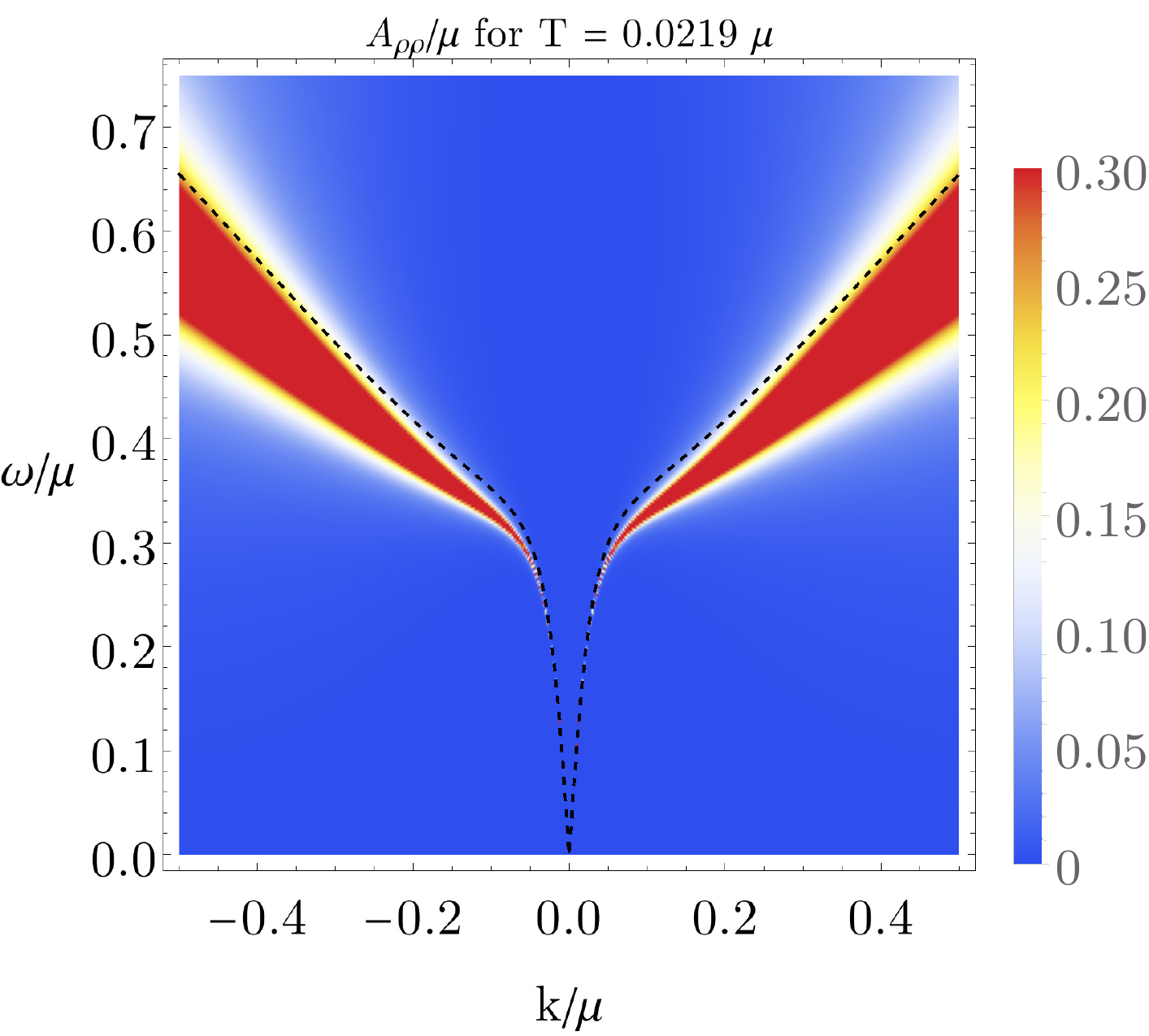}
      \caption{$p \ell = \pi/10$}
      \label{fig:layered_cos095}
    \end{subfigure}
    \vskip\baselineskip
     \begin{subfigure}{.5\textwidth}
        \centering
        \includegraphics[width=\linewidth]{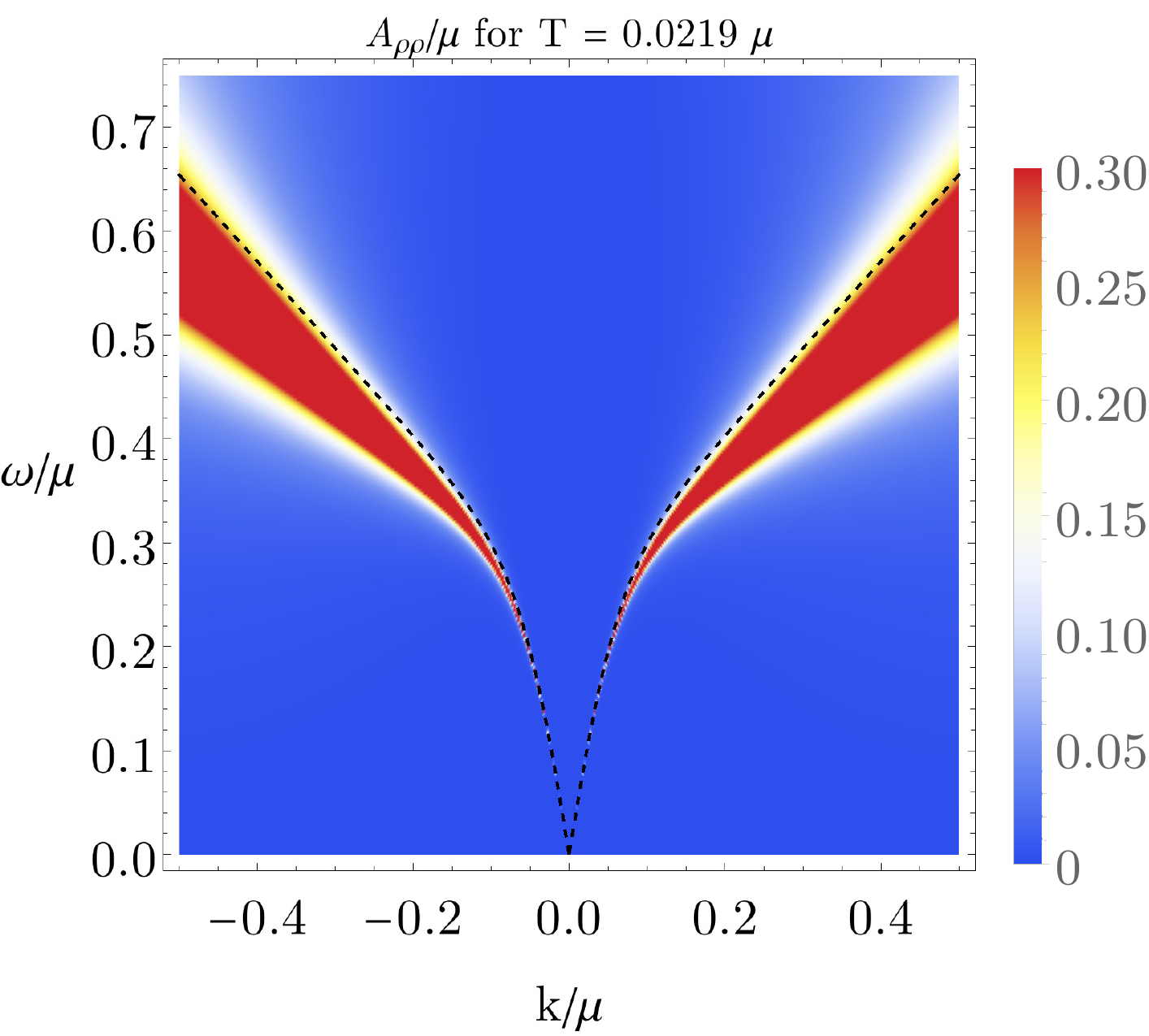}
        \caption{$p\ell = \pi /4$}
        \label{fig:layered_cos_pi4}
      \end{subfigure}%
    \begin{subfigure}{.5\textwidth}
      \centering
      \includegraphics[width=\linewidth]{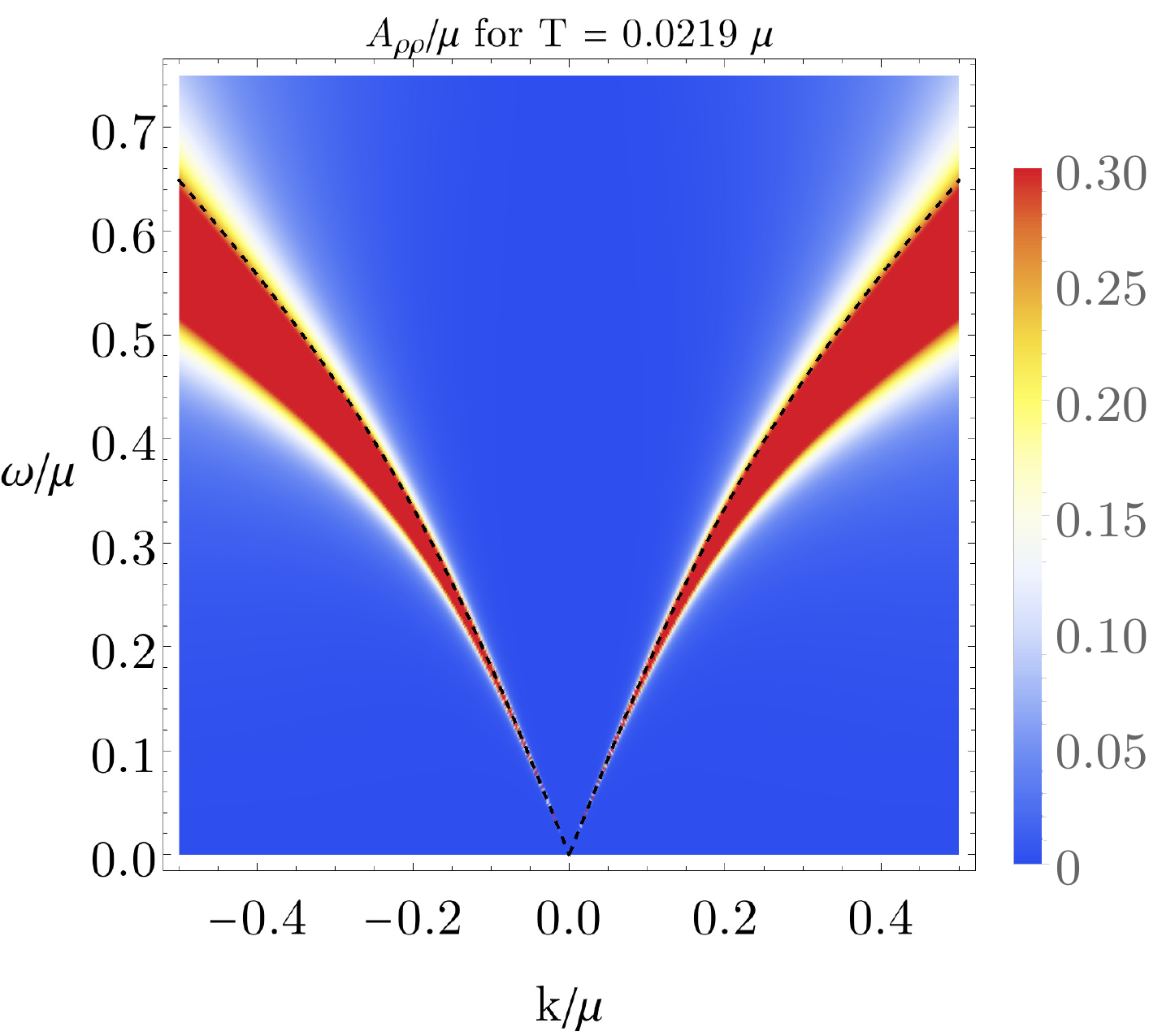}
      \caption{$p\ell = \pi$}
      \label{fig:layered_cos_pi}
    \end{subfigure}
    \caption{The density spectral function for a layered system at $T \simeq 0.02 \mu$. When the sources are in phase (a), i.e., $\cos(p\ell) = 1$, we recover the gapped mode of $(3+1)$-dimensional materials. However, if the sources are out of phase, for $k\ell \ll \abs{\cos(p\ell)}$ the dispersion relation is linear. We can also observe that for $\abs*{\cos(p\ell)} \ll k\ell \ll 1$, we recover the quadratic modes (b). For $k\ell \gg 1$ we instead recover the dispersion relation $\omega \propto \sqrt{\abs{\bm k}}$, as more noticeable in (c)-(d). The black dashed line represent the plasmon modes computed in the hydrodynamic approximation. In all the plots we used $\alpha^2 = 1$ and $\ell = 10$. }
   \label{fig:layered}
   \end{figure}

\begin{figure}\centering
    \includegraphics[width=0.6\textwidth]{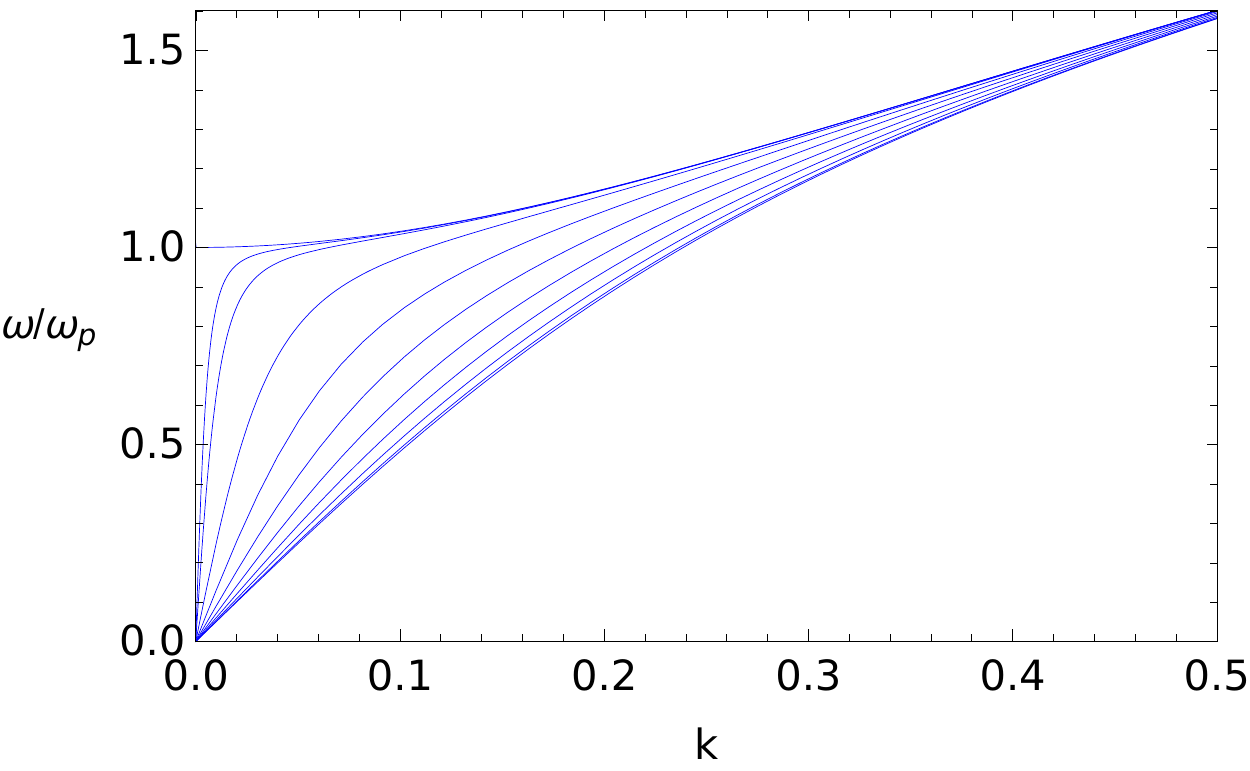}
     \caption{Effect of the double-trace deformation for a layered system on a response function with low-energy sound modes. For $p\ell = 0$ we obtain a gapped plasmon mode, while for any value of $p\ell \in (0, \pi]$ the low-energy dispersion relation is linear, with $p\ell = \pi$ having the lowest speed. The plot shows the plasmon dispersion for  $p\ell = \{0, \pi/50, \pi/25, n\pi/8\}$, with $n \in [1,8]$ an integer.}
     \label{fig:hydro_layered}
  \end{figure}

    Using eq.\ \eqref{eq:current_stack} into the boundary deformation in eq.\ \eqref{eq:maxwell_boundary} and choosing the gauge $\xi = 1$ for convenience, we obtain the action
    \begin{align}
      \int \dif^{d} x\, \int \dif z\, \left(\frac{1}{2 \alpha^2} A_\mu (x, z)\left[\eta^{\mu\nu}\partial^2 \right]A_\nu(x, z) - \sum_n A_\mu(x, z) \avg{J^\mu(x, z)} \delta(z - n\ell) \right) \text{ ,}
    \end{align}
    with $\mu = 0, \dots, d - 1$, and we have integrated out $A_z(x, z)$. Fourier transforming and integrating out the Maxwell field, we obtain
    \begin{align}
      -\frac{\alpha^2}{2}\int \frac{\dif^{d} k}{(2 \pi)^d} \int \frac{\dif k_z }{2 \pi} \left( \sum_{n,m} \avg{J^\mu(-k, n\ell)}\eta_{\mu\nu} \frac{e^{-i k_z(n - m) \ell}}{k^2 + k_z^2} \avg{J^\nu(k , m\ell)} \right) \text{ .}
    \end{align}
    We can next perform the integral over the momentum $k_z$ and further Fourier transform $$J^\mu(k, n\ell) = \frac{\ell}{2 \pi} \int_{-\pi/\ell}^{\pi/\ell} \dif p \, J^\mu(k, p) e^{i p n\ell} \text{ ,}$$
    so that we can then sum over the layer indices to obtain the desired double-trace deformation (see appendix \ref{app:layerd} for the detailed computation)
    \begin{align}\label{eq:layered_double_trace}
      -\frac{\alpha^2}{2}\int \frac{\dif^{d} k}{(2 \pi)^d} \int_{-\pi/\ell}^{\pi/\ell} \frac{\dif p}{2\pi}\, \avg{J^\mu(-k, -p)} \eta_{\mu\nu} \frac{\ell}{2\abs*{k}} \frac{\sinh{(\abs*{k} \ell)}}{\cosh{(\abs*{k} \ell)} - \cos(p \ell)} \avg{J^\nu(k , p)} \text{ ,}
    \end{align}  
    that gives the Coulomb potential for a layered electron gas.
    Notice that when the layer sources are in phase, i.e., $\cos(p\ell) = 1$, we recover the potential for a $(3+1)$-dimensional system proportional to $1/k^2$ for $\abs*{k} \ll 1$ and $\ell$ such that $\abs*{k} \ell \ll 1$, since
    \begin{align}
      \frac{\ell}{2\abs*{k}} \frac{\sinh{(\abs*{k} \ell)}}{\cosh{(\abs*{k} \ell)} - 1} = \frac{1}{k^2}  + \mathcal{O}(\ell^2) \text{ .} 
    \end{align}
    The spectral function then contains, for $p\ell = 0$, a gapped plasmon mode, as it is shown in figure \ref{fig:layered_cos_2pi}, where we plot the density spectral function at $T\simeq 0.02 \mu$ for a Reissner-Nordstr\"om solution with a double-trace deformation of the form of eq.\ \eqref{eq:layered_double_trace}. However, as the only coupling between the layers is due to the Coulomb interaction, the Green's functions of each single layer without the double-trace deformation contain linear modes typical of a $(2+1)$-dimensional system and with a dispersion $\omega = \abs{\bm k}/\sqrt{2}$ for low energies, and the resulting gapped mode is therefore different from the one in a $(3+1)$-dimensional material with the same characteristics.
    On the other hand for any $\cos(p \ell)\ne 0$ we can distinguish three different regimes. For $\abs{k}\ell \ll \abs{\cos(p\ell)}$ we obtain a constant potential
    \begin{align}
      \frac{\ell}{2\abs*{k}} \frac{\sinh{(\abs*{k} \ell)}}{\cosh{(\abs*{k} \ell)} - \cos(p\ell)} = \frac{\ell^2/2}{1 - \cos(p\ell)} + \mathcal{O}\left(\abs*{k}^2 \ell^4\right) \text{ .} 
    \end{align}
    In the Reissner-Nordstr\"om system with a linear low-energy dispersion relation, this has the effect of renormalizing the speed of these sound modes.
    However, for $\abs{\cos(p\ell)} \ll \abs{k}\ell \ll 1$, we recover the $1/k^2$ potential. Finally, for $\abs{k} \ell \gg 1$ we obtain a $\ell/2\abs{k}$ potential. As for the $(2+1)$-dimensional case, in condensed-matter systems we are interested in the limit where $\omega$ in the potential satisfies $\omega \ll \abs{\bm k}$ as it is suppressed by a factor of $v_\mathsc{f}/c \ll 1$, with $v_\mathsc{f}$ the Fermi velocity. In this limit, the latter potential gives rise to a dispersion relation $\omega \propto \sqrt{\abs{\bm k}}$. This behavior can be seen in figures \ref{fig:layered_cos095}-\ref{fig:layered_cos_pi}. In particular, by looking at the hydrodynamic approximation 
    \begin{align}
      G^{00}(\omega, \bm k) \simeq \frac{2\avg{\rho}^2}{3 \avg{\epsilon}}\frac{\bm k^2/\ell}{\omega^2 - \bm k^2 v_s^2} \quad \text{ for } \quad \omega \gtrsim \bm k \text{ ,}
    \end{align}
    we can see that, by introducing a double-trace deformation as in eq.\ \eqref{eq:layered_double_trace}, we obtain a response function 
    \begin{align}
      \chi(\omega, \bm k) = \frac{\bm k^2/\ell}{\omega^2 - v_s^2\bm k^2 - \frac{\alpha^2}{2\abs{\bm k}} \left(\frac{2\avg{\rho}^2}{3 \avg{\epsilon}}\right) \frac{\sinh(\abs{\bm k} \ell)}{\cosh{(\abs{\bm k} \ell)} - cos{(p\ell)}} \bm k^2} \text{ ,}
    \end{align}
    whose poles are shown in figure \ref{fig:hydro_layered} for several value of $p\ell$, from $p\ell = 0$, that gives the gapped plasmon mode, to $p\ell = \pi$ that leads to the lowest speed of sound, since in the hydrodynamic limit the renormalized speed of sound is given by 
    \begin{align}
      \tilde v_s^2 = \frac{1}{2} + \frac{\alpha^2}{2}\left(\frac{2\avg{\rho}^2}{3 \avg{\epsilon}}\right) \frac{\ell}{1- \cos{p\ell}} \text{ .}
    \end{align}
    In figures \ref{fig:layered_cos_2pi}-\ref{fig:layered_cos_pi} we see that the low-energy behavior of the acoustic plasmon modes is well described by the hydrodynamics approximation (shown as a black dashed line). As we move to higher frequencies, however, we observe a discrepancy between the holographic result and the hydrodynamic approximation. This is not unexpected as hydrodynamics is only a long-wavelength theory.

  \section{Plasmons in $3+1$ dimensions and conductivity}\label{sec:3dplasmon}
    In this section we apply the double-trace deformation to the holographic calculation of the $(3+1)$-dimensional optical conductivity in the Reissner-Nordstr\"om metal. Although it would seem straightforward to extend the results of the previous section to a $(3+1)$-dimensional model, there is an important subtlety as, in all even boundary dimensions, the theory contains a logarithmic divergence that introduces a scale into the theory that needs to be determined experimentally. Without a double-trace deformation this scale only enters the real part of the Green's functions, and it, therefore, does not modify the spectral functions. However, when introducing the double-trace deformation the scale affects the imaginary part of the Green's functions as well, and can be observed in the spectral function, or equivalently, in the real part of the (longitudinal) conductivity. We show that this gives rise to a form of the conductivity that resembles the one measured in Weyl and Dirac semimetals.

    \subsection{Renormalization and anomaly}
      In order to study the properties of the boundary theory we need to regularize the divergences of the boundary action. The necessary counterterms depend on the spacetime dimension of the theory, and for the gravitational part of the action they are treated in detail in ref.\ \cite{DeHaro2001} up to $d = 6$. 

      In the case of $d = 2 + 1$ treated above, the Maxwell term is finite as $r \to \infty$, and the only UV divergence comes from the gravitational part that we regularized by adding the counterterm in eq.\ \eqref{eq:counterterm}. 
      More interesting, however, is the case of even boundary spacetime dimension, as the $3 + 1$-dimensional system we are interested in. The asymptotic expansion of the fields as $r \to \infty$ is 
      \begin{align}\label{eq:asymptotic_exp}
        \begin{split}
          h_{\mu\nu} \rightarrow& r^2\left(h^{(0)}_{\mu\nu} + h^{(2)}_{\mu\nu} r^{-2} + \dots + h^{(d)}_{\mu\nu} r^{-d} + \textrm{h}_{\mu\nu} r^{-d} \log{(r/\abs*{k})} + \mathcal O(r^{-d - 2} \log{r})\right)\\
          a_{\mu} \rightarrow& a_\mu^{(0)} + a_\mu^{(2)} r^{-2} + \dots + a_\mu^{(d - 2)} r^{-d + 2} + \mathrm{a_\mu} r^{-d + 2} \log{(r/\abs*{k}) + \mathcal O (r^{-d} \log{r})} \text{ ,}
      \end{split}
      \end{align}
      where the expansion coefficients are functions of $k$, and the logarithmic terms only appear in even dimensions. All the higher-order coefficients as well as the coefficient of the logarithmic term are local functions of the leading-order coefficients $a_\mu^{(0)}$ and $h_{\mu\nu}^{(0)}$ and can be determined by a near-boundary analysis. The exceptions are the coefficients related to the field theory response, $h^{(d)}_{\mu\nu}$ and $a_\mu^{(d-2)}$ in the present case, that require the full solution of the linearized equations of motion.\footnote{In the asymptotic expansion in eq.\ \eqref{eq:asymptotic_exp} it is usually convention to include all the momentum dependence in the coefficients, hence to  adsorb the $\log(1/\abs*{k})$ term into the coefficient with the same power of $r$. However, here we keep the momentum dependence in the term $\log(r/\abs*{k})$, as it makes the following argument more clear.} In $d = 3 + 1$, we see that in addition to the divergence regulated by the counterterm in \eqref{eq:counterterm}, both the Maxwell and the gravitational field then present a logarithmic divergence, related to the conformal anomaly of the boundary theory \cite{Skenderis2002}. 

      As we show below, the regularization of the logarithm introduce a renormalization scale,  $k_\mathsc{uv}$, that cannot be determined from the theory but enters as an experimental parameter. 
      Since we are mostly interested in studying the density-density response function, here we focus on the logarithmic term coming from the expansion of the Maxwell field. The case of the metric field is analogous.
      The boundary term relevant for the $\avg{J^\mu J^\nu}$ correlation functions is
      \begin{align}
        \delta S^{(2)}_{A,bdy} = -\lim_{r_\mathsc{uv} \to \infty}  \frac{1}{2}\int_{r = r_\mathsc{uv}} \dif^d x \, \sqrt{- \gamma}  n^\mu \gamma^{\nu\sigma} a_\sigma \partial_\mu a_\nu \text{ .}
      \end{align}
       Inserting the asymptotic expansion of eq.\ \eqref{eq:asymptotic_exp} in $d = 3 + 1$, and the expressions for the induced boundary metric and the vector normal to the boundary, we obtain
      \begin{align}
        \delta S^{(2)}_{A,bdy} = -\lim_{r_\mathsc{uv} \to \infty}  \frac{1}{2}\int_{r = r_\mathsc{uv}} \dif^4 x \, \eta^{\nu \sigma} a_\sigma^{(0)} \left(- 2 a_\nu^{(2)} + \mathrm a_\nu - 2 \mathrm a_\nu \log{(r/\abs*{k})} \right) \text{ .}
      \end{align}
      So, in order to take the limit we need to regularize the logarithmic divergence. This can be done by inserting a scale-dependent counterterm 
      \begin{align}\label{eq:log_counterterm}
        S_{c.t.} = -\frac{\log(r_\mathsc{uv}/\tilde{k}_\mathsc{uv})}{4}\int_{r = r_\mathsc{uv}} \dif^d x\, \sqrt{-\gamma} F_{\mu\nu}F^{\mu\nu}
      \end{align} 
      that gives us the regularized boundary term
      \begin{align}
        \delta S^{(2)}_{A,bdy} = \frac{1}{2}\int \dif^4 x \, \eta^{\nu \sigma} a_\sigma^{(0)} 2 \left[ a_\nu^{(2)} + \mathrm a_\nu \log\left(\frac{k_\mathsc{uv} }{\abs{ k } }\right) \right] \text{ ,}
      \end{align}
      where we defined $k_\mathsc{uv} = \tilde{k}_\mathsc{uv} \sqrt{e}$. We, therefore, see that the response of the current operator to small fluctuations depends on the choice of scale:
      \begin{align}
        \delta \avg{J^\mu} = \frac{\delta S^{(2)}_{A,bdy}}{\delta a_\mu^{(0)}} = 2 \eta^{\mu \nu} \left[ a_\nu^{(2)} + \mathrm a_\nu \log\left(\frac{k_\mathsc{uv} }{\abs{ k } }\right) \right] \text{ .}
      \end{align}
      The coefficients $\mathrm a_\nu$ can be expressed in terms of the source term by matching coefficients in a near-boundary expansion, and we obtain
      \begin{align}
        \begin{split}
          \mathrm a_0 =& -\frac{\abs{\bm k}}{2}(\omega a_x^{(0)} + \abs{\bm k} a_0^{(0)}) \text{ ,}\\
          \mathrm a_x =& \frac{\omega}{2}(\omega a_x^{(0)} + \abs{\bm k} a_0^{(0)}) \text{ .}
        \end{split}
      \end{align}
      This implies that the dependence of the Maxwell Green's functions on the renormalization scale $k_\mathsc{uv}$ appears in a purely real term
      \begin{align}
        G^{\mu\nu} = \widetilde{G}^{\mu\nu} - k^\mu k^\nu \log\left(\frac{k_\mathsc{uv} }{\abs{ k } }\right) \text{ ,}
      \end{align}
      were we denoted with $\widetilde{G}^{\mu\nu}$ the part of the Green's functions independent of $k_\mathsc{uv}$. 
      The logarithmic correction is thus not visible in the spectral function, nor in the real part of the conductivity. 
      However, due to the form of the RPA-like response function in eq.\ \eqref{eq:rpa}, this is not the case anymore when we introduce the double-trace deformation, and the scale-dependent term can also be observed in the real part of the optical conductivity
      \begin{align}\label{eq:conductivity_rpa}
        \sigma(\omega) = \frac{\chi^{xx}(\omega, \bm k = \bm 0)}{-i\omega} \text{ .}
      \end{align}
      In particular for $\bm k = \bm 0$, the $a_x$ fluctuations decouples from all the others, and the current-current Green's function takes the form 
      \begin{align}
        G^{xx}(\omega, \bm k = \bm 0) = -\frac{2 a_x^{(2)}(\omega)}{a_x^{(0)}(\omega)} - \omega^2 \log\left(\frac{\omega_\mathsc{uv}}{\abs{\omega}}\right) \equiv \widetilde{G}^{xx}(\omega, \bm k = \bm 0) - \omega^2 \log\left(\frac{\omega_\mathsc{uv}}{\abs{\omega}}\right) \text{ ,}
      \end{align}
      and, as mentioned previously, for the neutral system the scale-dependent logarithm only affects the imaginary part of the optical conductivity
      \begin{align}
        \sigma_0(\omega) = \frac{\widetilde G^{xx}(\omega, \bm k = \bm 0)}{-i \omega} - i \omega  \log\left(\frac{\omega_\mathsc{uv} } {\abs{ \omega } }\right) \equiv \widetilde \sigma_0 - i \omega  \log\left(\frac{ \omega_\mathsc{uv} } {\abs{ \omega } }\right)  \text{ .}
      \end{align}
      When we include the double-trace deformation we see from eq.\ \eqref{eq:response_double_trace}
      that the optical conductivity for a system with Coulomb interactions becomes
      \begin{align}\label{eq:conductivity_coulomb}
        \sigma(\omega) = \frac{\sigma_0(\omega)}{1 - \frac{\alpha^2}{i\omega} \sigma_0(\omega)} \text{ ,}
      \end{align}
      so that the real part of the conductivity now contains a signature of the scale-dependent logarithm
      \begin{align}\label{eq:re_conductivity_coulomb}
        \text{Re}[\sigma(\omega)] = \frac{\text{Re}[\widetilde \sigma_0(\omega)]}{\left(1 - \frac{\alpha^2}{\omega} \text{Im}[\widetilde \sigma_0(\omega)] + \alpha^2 \log\left(\omega_\mathsc{uv}/\abs{\omega}\right)\right)^2 + \frac{\alpha^2}{\omega} \text{Re}[\widetilde \sigma_0(\omega)]} \text{ .}
      \end{align}
      This form of the conductivity, with a logarithmic term depending on a scale $\omega_\mathsc{uv}$ to be determined experimentally, resembles the optical conductivity of Dirac and Weyl semimetals \cite{Roy2017}. In the next sections, we give some explicit examples of conductivities in $3 + 1$ dimensions with the double-trace correction. In particular, in section \ref{sec:solvable} we compute the optical conductivity in pure $\ads_5$, without backreaction, that can be found analytically. In section \ref{sec:3d_cond_RN} we instead study the effects of the double-trace deformation on the optical conductivity of the Reissner-Nordstr\"om metal.

  \subsection{Exact solution in $\ads_5$ background}\label{sec:solvable}
    In the simple case of a Maxwell action on a $\ads_5$ background without backreaction, which is the probe limit where the coupling constant $\lambda \to \infty$, we can compute the optical conductivity analytically. 
    We use this example to show the importance of the logarithmic term in the optical conductivity when adding a double-trace deformation. 
    In particular, we find that the optical conductivity in the neutral theory matches the results for a non-interacting Weyl or Dirac semimetal obtained from condensed-matter calculations \cite{Jacobs2014,Vivian2015}. Inserting the deformation then gives us the optical conductivity expected from Weyl and Dirac semimetals with only long-range Coulomb interactions in the RPA approximation. 

    Since we consider the limit of no backreaction on the metric, we are only interested in the Maxwell action 
    \begin{align}\label{eq:maxwell_ads}
      S = -\frac{1}{4} \int \dif^{d + 1} \sqrt{- g} F_{\mu\nu} F^{\mu\nu} \text{ ,}
    \end{align}
    with the AdS background metric (we use the variable $\rho \equiv 1/r$ for notational convenience)
    \begin{align}
      \dif s^2 = \frac{1}{\rho^2} \left(\dif \rho^2 + \eta_{\mu\nu} d x^\mu d x^\nu\right) \text{ .}
    \end{align}
    The linearized equations of motion of the theory are solved, after fixing the momentum in the $x$ direction, by  
    \begin{align}
      a_x(\omega, \bm k) = 
      \begin{cases}
        C_1 J_{d/2 - 1}(-i k \rho) + C_2 Y_{d/2 - 1}(-i k \rho), & -\omega^2 + \bm k^2 > 0 \\
        \tilde C_1 H^{(1)}_{d/2 - 1}(\tilde{k} \rho) + \tilde C_2 H^{(2)}_{d/2 - 1}(\tilde{k} \rho), & -\omega^2 + \bm k^2 < 0     
      \end{cases}
      \text{ ,}
    \end{align}
    with $\tilde{k} \equiv \sqrt{-(-\omega^2 + \bm k^2)} \in \mathbb R$, and $H^{(1)}$ and $H^{(2)}$ are the Hankel functions of the first and second kind, respectively. From now on we focus on the case $- \omega^2 + \bm k^2 < 0$ as we are ultimately interested in the conductivity for $\bm k = \bm 0$. The solution with $\tilde C_2 = 0$ correspond to the infalling boundary condition and can be used to compute the retarded Green's function. We thus set $\tilde C_2 = 0$ in the following. 

    For $d = 3 + 1$, we can then study the near boundary ($\rho \to 0$) behavior of the field fluctuations
    \begin{align}
      a_x = \tilde C_{1} \left(-\frac{2 i}{\tilde k \pi} + \frac{\tilde k \rho^2}{2} + \frac{i \tilde k(2 \gamma_\mathsc{e} - 1)}{2\pi} \rho^2 + \frac{i \tilde k}{\pi} \log{\left(\frac{\tilde{ k}\rho}{2} \right)} \rho^2 \right)
    \end{align} 
    that shows the logarithmic term that gives rise to a divergence on the boundary, with $\gamma_\mathsc{e} \simeq 0.577$ the Euler's gamma constant. After regularizing the boundary action with a counterterm of the form 
    \begin{align}
      S_{c.t.} = -\frac{\log\left(\rho_\mathsc{uv} k_\mathsc{uv} e^{(2\gamma_\mathsc{e} - 1)/2}/2\right)}{4}\int_{\rho = \rho_\mathsc{uv}} \dif^d x\, \sqrt{-\gamma} F_{\mu\nu}F^{\mu\nu}
    \end{align}
    we can extract the optical conductivity
    \begin{align}\label{eq:ads_conductivity}
      \sigma_0(\omega) = \frac{G^{xx}(\omega, \bm k = \bm 0)}{-i \omega} = \frac{\pi}{2}\omega + i \omega \log{\left(\frac{\abs*{\omega}}{\omega_\mathsc{uv}}\right)}
      = \frac{i \omega}{2} \log\left(- \left( \frac{\omega}{\omega_\mathsc{uv}}\right)^2\right) 
      \text{ ,}
    \end{align}
    that exactly reproduces the zero-temperature optical conductivity for a non-interacting Weyl or Dirac semimetal, up to a prefactor \cite{Jacobs2014}. In particular, the holographic result in \eqref{eq:ads_conductivity} describes a system with strong interactions, and, contrary to the non-interacting Dirac semimetal calculation, the prefactor that for convenience we have set to one by a rescaling of the action, depends on the coupling constant. The frequency dependence, however, is the same as it is set by the conformal symmetry of the theory. 

    When introducing a double-trace deformation the new response function takes the form
    \begin{align}
      \chi^{xx}(\omega, \bm k) = \frac{G^{xx}(\omega, \bm k)}{1 - \frac{\alpha^2}{\omega^2} G^{xx}(\omega, \bm k)} \text{ ,}
    \end{align}
    so that the real part of the conductivity in the modified boundary theory with Coulomb interaction becomes:
    \begin{align}\label{eq:AdS_conductivity_RPA}
      \mathrm{Re}[\sigma(\omega)] = \mathrm{Re}\left[\frac{\chi^{xx}(\omega, \bm k = \bm 0)}{
        -i \omega} \right] = \frac{\frac{\pi}{2}\omega}{\left(1 + \alpha^2 \log(\abs*{\omega}/\omega_\mathsc{uv})\right)^2 + \frac{\alpha^4\pi^2}{4}} \text{ .}
    \end{align} 
    The logarithmic term in the denominator of the conductivity is reminiscent of the conductivity of Weyl semimetals \cite{Roy2017}. From \eqref{eq:AdS_conductivity_RPA}, we can see that for $\omega \to 0$ and $\omega \to \infty$, $\sigma \sim \omega/\log^2{(\omega)}$, and in both limits we have $\text{Re}[\sigma] < \text{Re}[\sigma_0]$. However, there is a range of values for the coupling constant $\alpha^2$ where we can have $\text{Re}[\sigma] > \text{Re}[\sigma_0]$ (considering only $\omega > 0$, where the real part of the conductivity is always positive). Explicitly:
    \begin{align}
      \begin{split}
      &\alpha^2 < \frac{2}{\pi}: \, \begin{cases}
        \sigma(\omega) > \sigma_0(\omega), & \exp{\left(-\frac{1}{\alpha^2} - \sqrt{\frac{1}{\alpha^4} - \frac{\pi^2}{4}}\right)} < \frac{\omega}{\omega_\mathsc{uv}} < \exp{\left(-\frac{1}{\alpha^2} + \sqrt{\frac{1}{\alpha^4} - \frac{\pi^2}{4}}\right)} \text{ ,}\\
        \sigma(\omega) < \sigma_0(\omega), & \text{otherwise}
      \end{cases}\\
      &\alpha^2 > \frac{2}{\pi}: \quad \sigma(\omega) < \sigma_0(\omega)\text{,} \quad \forall \omega \in \mathbb{R^+} \text{ .}
    \end{split}
    \end{align}

\begin{figure}[h]\centering
      \includegraphics[width=0.6\textwidth]{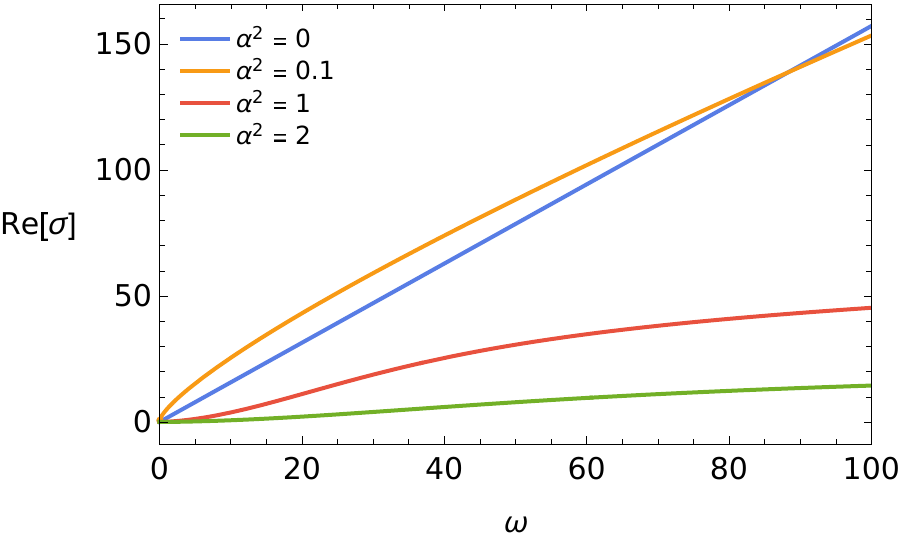}
       \caption{Conductivity for a Maxwell field in $\ads_5$ without backreaction, but with a double-trace boundary deformation. The value of the coupling constant $\alpha^2$ determines the importance of the effect of the double-trace deformation. The case $\alpha^2 = 0$ corresponds to the standard solution $\text{Re}[\sigma(\omega)] = \omega\pi/2$.}
       \label{fig:conductivity_rpa}
    \end{figure}

    This behavior can be observed in figure \ref{fig:conductivity_rpa}, were we plot the conductivity with and without double-trace deformation for different values of the coupling constant. In all the plots we fixed the renormalization scale $\omega_\mathsc{uv} = 100$. Note however that, being the only scale in the theory, a change in the value of $\omega_\mathsc{uv}$ simply amounts to a change of scale. 
    Notice also, that there is no plasmon peak in the conductivity. This is an effect of the probe limit, where the Maxwell field decouples from the gravitational background and does not backreact on the geometry. This in turns suppresses the sound modes in a neutral theory, and hence the plasmon modes in a charged system.
    
  \subsection{Conductivity with Coulomb interactions}\label{sec:3d_cond_RN}
  Here we compute the optical conductivity in $d = 3 + 1$ boundary spacetime dimensions for the Reissner-Nordstr\"om metal with and without double-trace deformation. 

  \begin{figure}\centering
    \includegraphics[width=.49\textwidth]{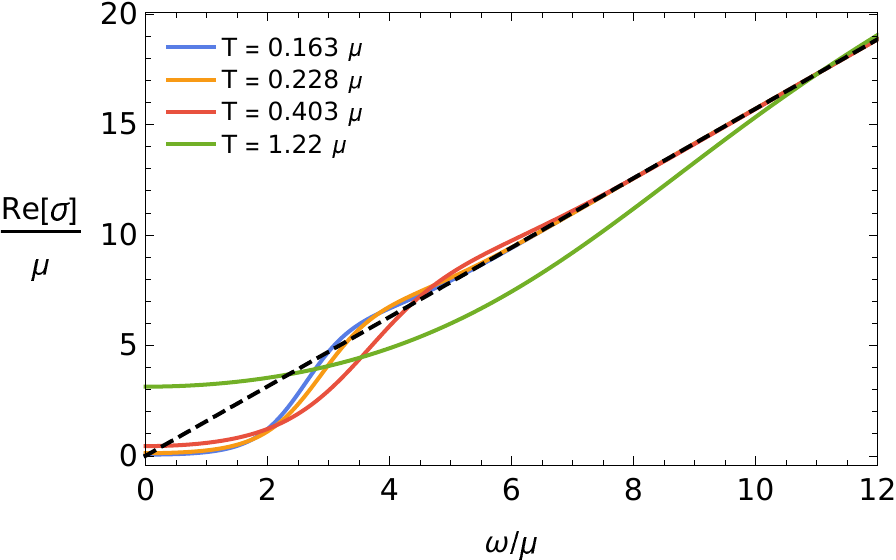}
    \includegraphics[width=.49\textwidth]{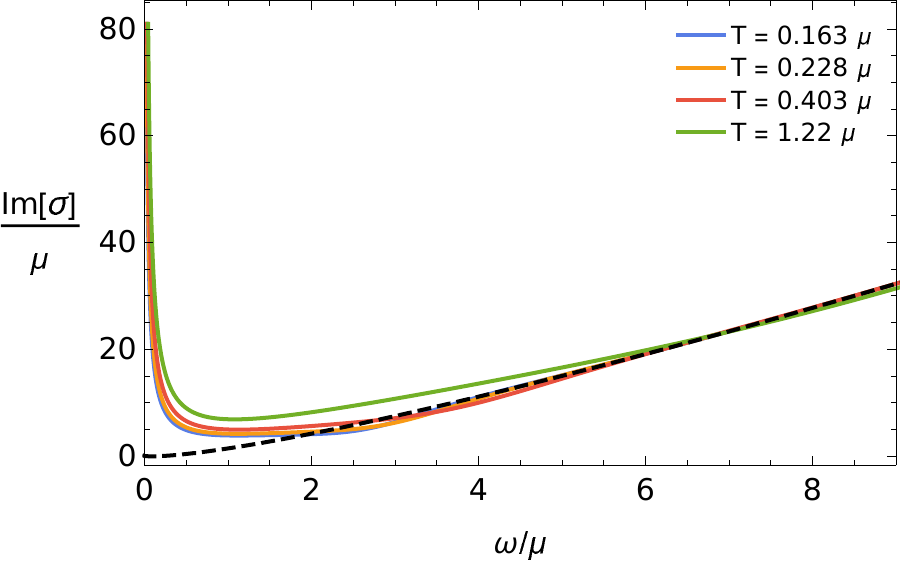}
  \caption{Real (left) and imaginary (right) part of the optical conductivity for the Reissner-Nordstr\"om metal in $d = 3 + 1$ for different values of the temperature. The black dashed line is the $\ads_5$ analytical solution. For large frequency we see that both the real and the imaginary parts go to the $\ads_5$ results, where for the imaginary part we choose the cutoff scale $\omega_\mathsc{uv}$ to be the same for different values of $T/\mu$. In particular, in the imaginary part, we can observe the logarithmic behavior for large frequency, as well as the $1/\omega$ behavior for small $\omega$. This implies that the real part of the conductivity contains a delta function $Z\delta(\omega)$, with $Z$ the Drude weight that in this case is given by $Z = 2 \avg{\rho}^2/3\avg{\epsilon}$.}
    \label{fig:RNconductivity_neutral}
\end{figure}
  
  In figure \ref{fig:RNconductivity_neutral}, we show the optical conductivity for a neutral system for different values of $T/\mu$ and compare it with the $\ads_5$ analytical solution obtained in the previous section. In particular, the short-wavelength limit of the theory is determined by the geometry of the bulk spacetime for large $r$, and we therefore expect to recover the $\ads_5$ result for high frequencies. This is indeed what we observe in figure \ref{fig:RNconductivity_neutral}, and we can notice the logarithmic behavior at large frequencies in the imaginary part of the conductivity, were we chose the value of the cutoff-scale $\omega_\mathsc{uv}$ to be the same for different values of $T/\mu$, since the asymptotic behavior depends on the choice of scale. Moreover, we can see that at small frequencies the imaginary part of the optical conductivity of the Reissner-Nordstr\"om solution diverges as $1/\omega$, signaling the presence of a delta function at the origin for $\text{Re}[\sigma(\omega)]$. This delta function is expected since we are dealing with a system with translation invariance, but it is suppressed in the probe limit for the $\ads_5$ solution as the background is kept fixed \cite{Horowitz2008}.

  \begin{figure}
  \centering 
  \includegraphics[width=.49\textwidth]{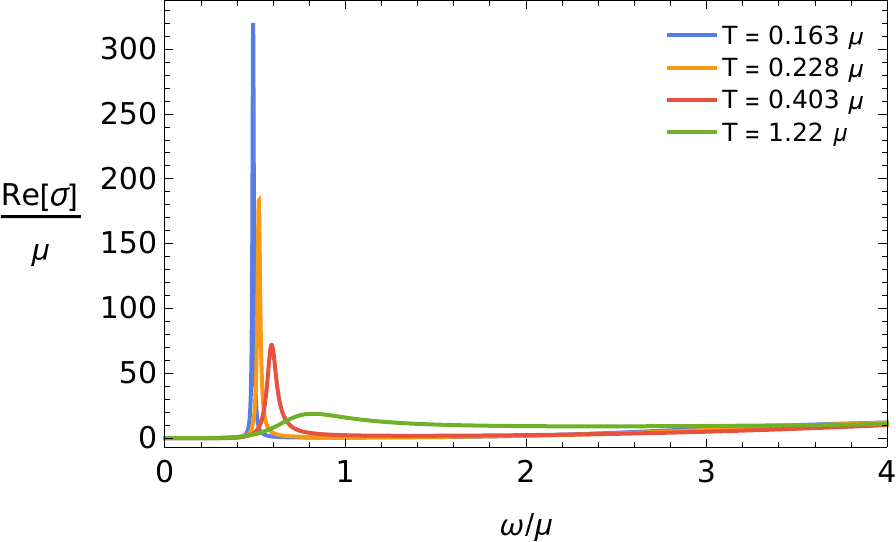}
  \hfill
  \includegraphics[width=.49\textwidth]{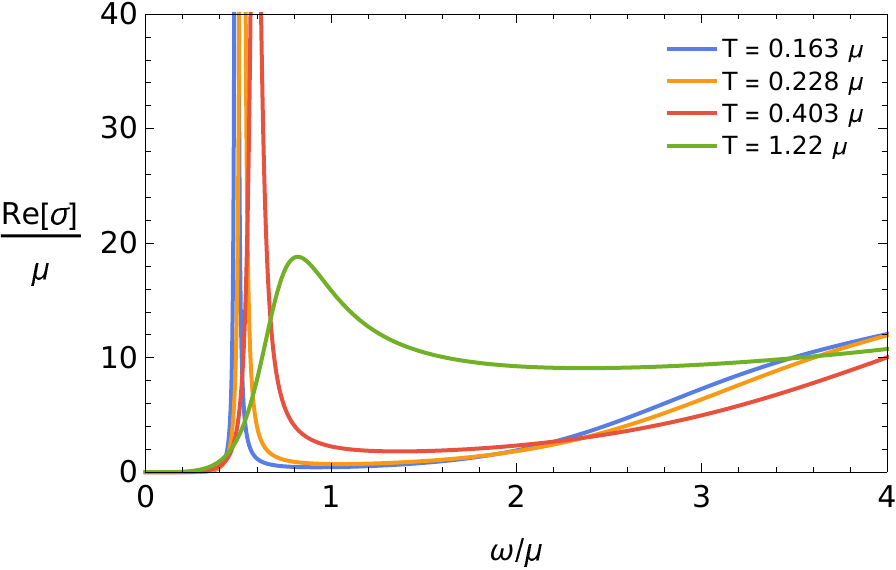}
  \caption{Real part of the optical conductivity for the Reissner-Nordstr\"om metal in $d = 3 + 1$ with a double-trace deformation ($\alpha^2 = 1/10$). On the right we show an enlarged version of the figure on the left. We clearly see a peak at the plasma frequency, that becomes increasingly higher (left) and sharper (right) as we lower the temperature. Moreover, we notice that the conductivity goes to zero as $\omega \to 0$, contrary to the neutral system.}
    \label{fig:RNconductivity_rpa}
\end{figure}

\begin{figure} 
\centering 
\includegraphics[width=.6\textwidth]{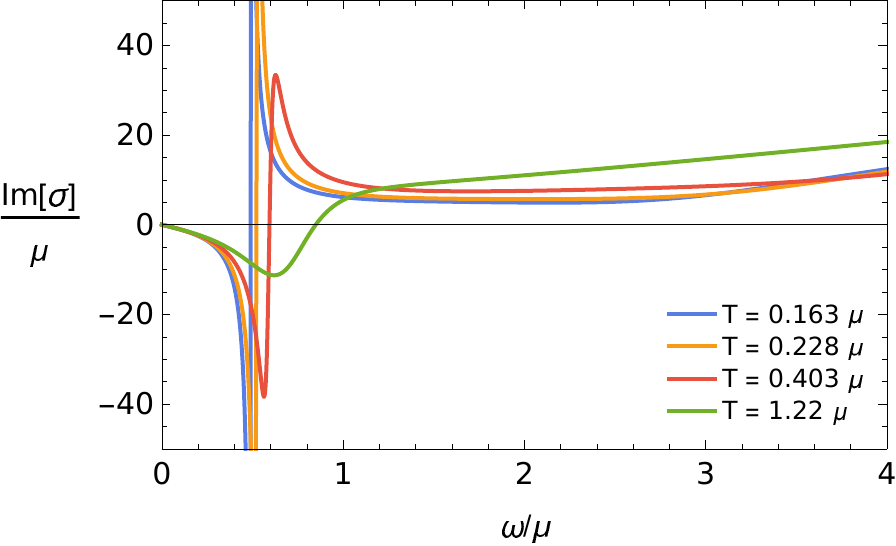}
\caption{Imaginary part of the optical conductivity for the Reissner-Nordstr\"om metal in $d = 3 + 1$ with a double-trace deformation ($\alpha^2 = 1/10$). We see that $\text{Im}[\sigma(\omega)] \to 0$ for $\omega \to 0$. This implies that there is no pole at the origin, as screening effects shift the pole to the nonzero plasma frequency.}
  \label{fig:RNconductivity_rpa_im}
\end{figure}

  By introducing the double-trace deformation the delta function in the origin disappears, as it is turned into a peak at the plasma frequency by Coulomb interactions. In figure \ref{fig:RNconductivity_rpa}, we can clearly observe this peak, that becomes sharper and higher as we lower the temperature. In figure \ref{fig:RNconductivity_rpa_im} we plot the imaginary part of the conductivity, where we see that the pole in the origin present in the optical conductivity of a neutral system is shifted to the nonzero plasma frequency. Moreover, with the double-trace deformation, the real part of the conductivity goes to zero as $\omega \to 0$, as is expected from eq.\ \eqref{eq:re_conductivity_coulomb}.

\section{Conclusion and Outlook}
  In summary, we proposed a general procedure to introduce screening effects of the Coulomb interaction in the holographic description of strongly interacting system. This allows us to study spectral functions of systems with charged particles that contain plasmon excitations, as it is the case in many condensed-matter systems of interest. In particular, we numerically studied the effect of this procedure in a Reissner-Nordstr\"om theory, where we observed properties expected from traditional condensed-matter calculations. 
  
  In $d = 3 + 1$ we obtained a gapped plasmon mode in the density-density response function, and we showed that the optical conductivity with Coulomb interactions contains scale-dependent logarithms, resembling the conductivities predicted in Dirac and Weyl semimetals.
  
  In $d = 2 + 1$, our main result is a toy model of a system composed of a stack of (spatially) two-dimensional layers, with strong in-plane interaction and with coupling between layers governed by the Coulomb interaction. In this model  we see that the behavior of the low-energy dynamics depends on the out-of-plane Bloch momentum $p$. When $p\ell = 0$, with $\ell$ the inter-plane distance the density-density spectral function contains a gapped plasmon mode. However, for $p\ell \ne 0$, we obtain an `acoustic plasmon', with a linear low-energy dispersion relation with a renormalized speed of sound.
  These results show that the model qualitatively reproduces recent experimental results \cite{Hepting2018}. This suggests that the Coulomb interaction between layers might play a key role in high-temperature cuprate superconductors, proving the necessity of incorporating this interaction in holographic models if we want to study properties of these layered high-$T_c$ materials. In a single layer, we instead showed that the dispersion relation assume the form $\omega \propto \sqrt{\abs{\bm k}}$, as observed in graphene. 

  In conclusion, we have seen that, even with a relatively simple holographic model as the Reissner-Nordstr\"om background, the addition of long-range Coulomb interactions presents interesting features that can more closely reproduce experimental results in strongly inte\-ract\-ing materials. Therefore, it would be very interesting in future work to apply the double-trace deformation introduced here to different holographic backgrounds, such as the holographic superconductor model and Lifshitz solutions, in order to study the behavior of their longitudinal low-energy excitations in the presence of Coulomb interactions. These gravitational theories, contrary to the Reissner-Nordstr\"om model,  allow for the description of systems with zero entropy at zero temperature. As the ultimate goal is to describe laboratory condensed-matter system, a next important step is to relax the assumption of momentum conservation, as impurities in experimental materials necessarily break momentum conservation. Finally, the addition of a hyperscaling-violation factor to the metric of the Lifshitz solution can be used to study the effect of Coulomb interactions on quantum phases with hyperscaling violation \cite{Krikun2018}.

\acknowledgments

We are grateful to Tobias Zingg and J\"org Schmalian for helpful suggestions and stimulating discussions. We would also like to thank Nick Plantz and Francisco Garc\'ia Florez for useful advice during the writing of this paper. This work is supported by the Stichting voor Fundamenteel 
Onderzoek der Materie (FOM) and is part of the D-ITP consortium, a
program of the Netherlands Organization for Scientific
Research (NWO) that is funded by the Dutch Ministry
of Education, Culture and Science (OCW).

\appendix

  \section{Gauge Solutions}\label{app:gauge_solutions}  
    In the theory considered in this paper, we have two gauge fields, the one-form $A_\mu$ and the metric tensor $g_{\mu\nu}$. In order to compute the Green's function, we then introduce fluctuations of these fields $\delta A_\mu \equiv a_\mu$ and $\delta g_{\mu\nu} \equiv h_{\mu\nu}$. As explained in section \ref{sec:reissner_nord}, using rotational invariance to fix the momentum along the $x$ direction $k^\mu = (\omega, \pm\abs{\bm k}, 0, \dots, 0)$, the remaining fluctuations decouple according to their parity under $O(d - 2)$ acting on $x^2, \dots, x^{d-1}$ \cite{Kovtun2005a}. In the longitudinal channel, in both $d = 3 + 1$ and $d = 2 + 1$, we are then left with (where obviously in $d = 2+ 1$ there is no $h_{zz}$)
    \begin{align}
      \delta \bm \Phi = (a_t, a_x, a_r, h_{xt}, h_{tt}, h_{xx}, h_{yy} = h_{zz}, h_{rr}, h_{tr}, h_{xr})\text{ .}
    \end{align}
    Using the gauge freedom to choose a gauge where $h_{r\mu} = 0$, as well as $A_r = a_r = 0$, the set of coupled fluctuations reduces to 
    \begin{align}
      \delta \bm \Phi = (a_t, a_x, h_{xt}, h_{tt}, h_{xx}, h_{yy} = h_{zz})\text{ .}
    \end{align}
    This set of fluctuations contains only two physical degrees of freedom, therefore, we can only find two independent solutions to the coupled system of linearized equations of motion. However, after setting the $r$ components to zero, we still have some left-over gauge freedom, and the remaining solutions necessary to compute the Green's functions are pure gauge solutions that can be extracted once we know the residual gauge freedom we are left with, as we explain below.

    The fields are invariant under the gauge transformation
    \begin{align}\label{eq:gauge_transformation}
      A_\mu + a_\mu \rightarrow A_\mu + a_\mu - \partial_\mu \Lambda \text{ ,}
    \end{align}
    and diffeomorphisms. The latter gives
    \begin{align}
      \begin{split}
      a_\mu &\rightarrow a_\mu -  \xi^\nu \nabla_\nu A_\mu - \left(\nabla_\mu \xi^\nu\right) A_\nu\\
      h_{\mu\nu} &\rightarrow h_{\mu\nu} - \nabla_\mu \xi_\nu - \nabla_\nu \xi_\mu \text{ .}
      \end{split}
    \end{align}

    In order to set the $r$-components of the metric fluctuations to zero we have to choose a
    vector $\bar \xi^\nu$ such that 
    \begin{align}\label{eq:gaugePDEs}
      h_{r\mu} =\nabla_r \bar\xi_\mu - \nabla_\mu \bar\xi_r = \partial_r \bar \xi_\mu - 2 \Gamma^{\lambda}_{r\mu} \bar \xi_\lambda + \partial_\mu \bar \xi_r \text{ .}
    \end{align}
    This defines a set of partial differential equations. Working in Fourier space and using rotational invariance to fix the momentum in the $x$ direction, we obtain a set of coupled ordinary differential equations of the form 
    \begin{align}
      \xi'(r,\omega, k) + A(r) \xi(r,\omega, k) = B(r, \omega, k) \text{ ,}
    \end{align}
    that has the general solution
    \begin{align}\label{eq:general_solution}
      \bar \xi(r, \omega, k) = \left(\int \mathrm d r \left[B(r, \omega, k) e^{\int \dif r A(r)}\right] + C(\omega, k)\right) e^{-\int \dif r A(r)} \text{ .}
    \end{align}
    We therefore see that $\xi_\mu$ is determined up to an arbitrary factor $C_\mu(\omega, k) e^{-\int \dif r A_\mu(r)}$, that corresponds to the left over gauge freedom. In particular from eq.\ \eqref{eq:gaugePDEs} we find that the residual gauge transformations are given by
    \begin{align}
      \xi_r =& C_r(\omega, k) e^{2 \int \dif r \Gamma^r_{rr}} \text{ ,}\\
      \xi_\alpha =& C_\alpha(\omega, k) e^{2 \int \dif r \Gamma^\alpha_{r\alpha}} \text{ ,}
    \end{align}
    where $\alpha = t, x, y$ and there is no summation on the two repeated indices in $\Gamma^\alpha_{r\alpha}$. 
    In the same way, we want to choose the scalar $\Lambda$ in \eqref{eq:gauge_transformation} in order to set the $r$ component of the gauge field to zero. Under the diffeomorphism used to set $h_{r\mu} = 0$, we have that $A_r + a_r$ transformed as
    \begin{align}
      A_r + a_r \rightarrow A_r + a_r + 2 g^{tt} \bar\xi_t \Gamma^t_{rt}A_t - g^{tt} \bar\xi'_t A_t \text{ ,}
    \end{align}
    so we need to choose $\Lambda$ to be
    \begin{align}
      \Lambda(r, \omega, k) = \int \dif r \left(A_r + a_r + 2 g^{tt} \bar\xi_t \Gamma^t_{rt}A_t - g^{tt} \bar\xi'_t A_t \right) + \lambda(\omega, k)
    \end{align}
    where $\lambda(\omega, k)$ is an arbitrary constant in $r$. Notice that this expression is also invariant under the residual gauge transformation $\bar\xi_t \rightarrow \bar\xi_t + c_t e^{2 \int \dif r \Gamma^{t}_{rt}}$. 
    Since we have some gauge freedom left, and we know that the linearized equations of motion are invariant under all allowed gauge transformation, we have, given a solution $\bar\Phi$ of the linearized equations of motion, that $\bar\Phi$ transformed under all residual gauge transformation is also a solution. Moreover, we know that $\Phi = 0$ is a solution. We can thus construct pure gauge solutions by choosing $n$ independent residual gauge transformation, i.e., by fixing $n$ linearly independent values of the vector $\bm C = (\lambda, C_\alpha)$.
    
    For example, in $d = 2 + 1$ we have from eqs. \eqref{eq:gaugePDEs} and \eqref{eq:general_solution} that the gauge where there are no $r$-components of the fluctuations is defined by
    \begin{align}\label{eq:gaugefix}
      \begin{split}
        \bar\xi_r =& \left(\frac{1}{2}\int \dif r\left[h_{rr} e^{-\int \dif r \Gamma^r_{rr}}\right] + C_r\right) e^{\int \dif r \Gamma^r_{rr}} \text{ ,}\\
        \bar\xi_t =& \left( \int \dif r\left[(h_{rt} + i\omega \xi_r) e^{-2 \int \dif r \Gamma^t_{rt}}\right] + C_t\right) e^{2 \int \dif r \Gamma^t_{rt}} \text{ ,}\\
        \bar\xi_x =& \left( \int \dif r\left[(h_{rx} - i k \xi_r) e^{-2 \int \dif r \Gamma^x_{rx}}\right] + C_x\right) e^{2 \int \dif r \Gamma^x_{rx}} \text{ ,}\\
        \bar\xi_y =& \left( \int \dif r\left[h_{ry} + e^{-2 \int \dif r \Gamma^y_{yt}}\right] +C_y\right) e^{2 \int \dif r \Gamma^y_{ry}} \text{ ,}
      \end{split}
    \end{align}
    and we are left with the gauge transformations defined by 
    \begin{align}\label{eq:residualgauge}
      \begin{split}
        \Lambda =& \lambda(\omega, k) \text{ ,}\\
        \xi_i =& C_i(\omega, k) e^{2 \int \dif r /r} = r^2 C_i(\omega, k) \text{ ,} \\
        \xi_t =& C_t(\omega, k) e^{\int \dif r \frac{f'}{f}} = f C_t(\omega, k) \text{ ,}\\
        \xi_r =& C_r(\omega, k) e^{-\int \dif r \frac{f'}{2f}} = \frac{1}{\sqrt{f}} C_r(\omega, k) \text{ .}
      \end{split}
    \end{align}
    Remembering that the behavior of the considered field fluctuations under a gauge transformation is given, in Fourier space, by
    \begin{align}\label{eq:gaugeBehavior}
      \begin{split}
        a_x \rightarrow& a_x - ik\Lambda \text{ ,}\\
        a_t \rightarrow& a_t + i\omega \Lambda \text{ ,}
      \end{split}
    \end{align}
    and
    \begin{align}\label{eq:diffBehavior}
      \begin{split}
        a_x \rightarrow& a_x + ik A_t/f \xi_t \text{ ,}\\
        a_t \rightarrow& a_t -f A'_t \xi_r - i\omega/f A_t \xi_t \text{ ,}\\
        h_{xt} \rightarrow& h_{xt} -i k \xi_t + i\omega \xi_x \text{ ,}\\
        h_{tt} \rightarrow& h_{tt} + 2 i \omega \xi_t +f f' \xi_r \text{ ,}\\ 
        h_{xx} \rightarrow& h_{xx} -2 ik \xi_x - 2f r\xi_r \text{ ,}\\
        h_{yy} \rightarrow& h_{yy} -2 f r\xi_r \text{ ,}
      \end{split}
    \end{align}
    we can generate 4 independent pure gauge solutions by setting the fluctuations to zero on the right-hand side of \eqref{eq:gaugeBehavior} and \eqref{eq:diffBehavior} and plugging in the expressions for the residual gauge freedom \eqref{eq:residualgauge} with $\bm C = C_\mu \delta^\mu_{\bar \nu}$ and $\bar\nu$ ranging over the 4 indices. 

    Notice that in \eqref{eq:gaugefix}, $\bar\xi_t$ and $\bar\xi_x$ depends on $\xi_r$, therefore, when performing a residual gauge transformation $\bar\xi_r \rightarrow \bar\xi_r + 1/\sqrt{f} C_r$, $\bar\xi_t$ and $\bar\xi_x$ will change as well. We ultimately find that the pure gauge solutions are:
    \begin{align}\label{eq:pureGaugeSolution}
      \begin{split}
        \delta\bm \Phi^{(1)} = &\left(-\omega, k, 0, 0, 0, 0\right)\\
        \delta\bm \Phi^{(2)} = &\left(0, 0, i\omega r^2, 0, -2 i k r^2, 0\right)\\
        \delta\bm \Phi^{(3)} = &\left(-i\omega A_t, i k A_t, -i k f, 2 i \omega f, 0, 0\right)\\
        \delta\bm \Phi^{(4)} = &\bigg(-\sqrt{f} A'_t + \omega^2 A_t \int \frac{\dif r}{f \sqrt{f}},
        -\omega k \int \frac{\dif r A_t}{f\sqrt{f}} + \omega k \int \frac{\dif r A_t}{f \sqrt{f}} , \\ &\omega k \left[f\int \frac{\dif r}{f\sqrt{f}} + r^2\int \frac{\dif r}{r^2\sqrt{f}}\right], -2\omega^2 f \int\frac{\dif r}{f \sqrt{f}} + \sqrt{f} f',\\&
         -2 k^2 r^2 \int\frac{\dif r}{r^2 \sqrt{f}} - 2 \sqrt{f} r, - 2\sqrt{f} r\bigg)
      \end{split}
    \end{align}
    This results can now easily be checked by substitution in the  linerarized equations of motion.

    \section{Details of the calculations for the layered system}\label{app:layerd}

    Using the notation of section \ref{subsec:layered}, with $d = 2 + 1$, the $(3 + 1)$-dimensional boundary deformation in Fourier space, after integrating out the Maxwell field, is, up to a factor of $-\alpha^2/2$ that we neglect here for notational convenience
    \begin{align}\label{eq:app_fourier}
          \int \frac{\dif^{d}  k}{(2 \pi)^{d} } \int \frac{\dif k_z}{2 \pi} \avg{J^{\mu}(- k, - k_z)} \frac{\eta_{\mu\nu}}{ k^2 + k_z^2} \avg{J^\nu( k , k_z)} \text{ ,}
    \end{align}
    with $J^\mu( k, k_z)$ the Fourier transform of $J^\mu( x, z)$ defined as 
    \begin{align}
      J^\mu( x, z) = \sum_n J^\mu( x, z) \delta(z - n \ell) \text{ .}
    \end{align} 
    The Fourier transform along the $z$ direction can easily be computed as
    \begin{align}
      \begin{split}
        J(x, k_z) =& \sum_n \int \dif z\, \int J(x, z) \delta(z - n\ell) e^{-i k_z z}
        = \sum_n J(x, n\ell) e^{-i k_z n \ell} \text{ .}
      \end{split}
    \end{align}
    Inserting this into \eqref{eq:app_fourier} we then have
    \begin{align}
      \int \frac{\dif^{d}  k}{(2 \pi)^{d} } \int \frac{\dif k_z}{2 \pi} \left( \sum_{n,m} \avg{J^{\mu}(- k, n\ell)}\eta_{\mu\nu} \frac{e^{-i k_z(n - m) \ell}}{ k^2 + k_z^2} \avg{J^\nu( k , m\ell)} \right) \text{ ,}
    \end{align}
    and we can perform the integral over $k_z$ that gives
    \begin{align}
      \int \frac{\dif^{d}  k}{(2 \pi)^{d}} \left( \sum_{n,m} \avg{J^\mu(- k, n\ell)} \eta_{\mu\nu} \frac{e^{-\abs*{n - m} \abs*{ k} \ell}}{2 \abs*{ k}} \avg{J^\nu( k , m\ell)} \right) \text{ .}
    \end{align}
    In order to perform the summation, we first Fourier transform $J(k, n\ell)$ in the first Brillouin zone, that is, we can write
    \begin{align}
      J(k, n \ell) =  \frac{\ell}{2 \pi} \int_{-\pi/\ell}^{\pi/\ell} \dif p \, J(k, p) e^{i p n \ell} \text{ ,}
    \end{align}
    to obtain 
    \begin{align}\label{eq:app_layered}
      \int \frac{\dif^{d}  k}{(2 \pi)^{d}}  \sum_{n,m} \frac{\ell^2}{(2 \pi)^2} \int_{-\pi/\ell}^{\pi/\ell} \dif p\, \dif p'\, \avg{J^\mu(- k, p)} \eta_{\mu\nu} \frac{e^{-\abs*{n - m} \abs*{ k} \ell}}{2 \abs*{ k}} \avg{J^\nu( k , p')} e^{i n\ell p + i m \ell p'}  \text{ .}
    \end{align}
    We can now perform the summations
    \begin{align}
      \begin{split}
      &\sum_{n, m} e^{-\abs*{n - m} \abs*{ k} \ell} e^{i n\ell p + i m \ell p'}\\
      & = \sum_m e^{-m \abs*{ k} \ell + i m \ell p'} \sum_{n < m} e^{n \abs*{ k} \ell + i n \ell p} + \sum_m e^{m \abs*{ k} \ell + i m \ell p'} \sum_{n > m} e^{-n \abs*{ k} \ell + i n \ell p} + \sum_m e^{i m\ell (p + p')}\\
      &= \sum_m e^{i m\ell (p + p')}\left(\sum_{n < 0} e^{n \abs*{ k} \ell + i n \ell p} + \sum_{n > 0} e^{-n \abs*{ k} \ell + i n \ell p} + 1 \right)\\
      &= \frac{2\pi}{\ell}\delta(p + p') \left(2 \sum_{n > 0} e^{-n \abs*{ k} \ell} \cos(p\ell) + 1\right) = \frac{2\pi}{\ell}\delta(p + p') \frac{\sinh{(\abs*{ k} \ell)}}{\cosh{(\abs*{ k} \ell)} - \cos(p \ell)} \text{ ,}
      \end{split} 
    \end{align}
    to finally obtain 
    \begin{align}
      \int \frac{\dif^{d}  k}{(2 \pi)^{d}} \int_{-\pi/\ell}^{\pi/ \ell} \frac{\dif p}{2\pi}\, \avg{J^\mu(- k, -p)} \eta_{\mu\nu} \frac{\ell}{2\abs*{ k}} \frac{\sinh{(\abs*{ k} \ell)}}{\cosh{(\abs*{ k} \ell)} - \cos(p \ell)} \avg{J^\nu( k , p)} \text{ .}
    \end{align}  

    For the boundary theory coming from holography, we use a set of $d$-dimensional solutions dual to a $d + 1$ bulk theory, whose boundary terms read, up to a factor of $1/2$
    \begin{align}
      \int \dif^d  x\,  A_\mu^{(n)}( x) \avg{J^\mu_{(n)}( x)} \text{ ,}
    \end{align}
    so that the $(d+1)$-dimensional stack of layers can be modelled as
    \begin{align}
      \sum_n \int \dif^d  x\,  A_\mu^{(n)}( x) \avg{J^\mu_{(n)}( x)} = 
        \int \dif^d x\,\int \dif z\, \sum_n  A_\mu( x, z) \avg{J^\mu( x, z)} \delta(z - nl)  \text{ .}
    \end{align}
    Fourier transforming as described above, we can rewrite this last integral as
    \begin{align}
      \begin{split}\label{eq:layered_bdy}
        &\int \frac{\dif^{d}  k}{(2 \pi)^{d}} \int \frac{\dif k_z}{2 \pi}\, \int_{-\pi/\ell}^{\pi/ \ell} \frac{\dif p}{2\pi}\, \ell A_\mu( k, k_z) \avg{J^\mu(- k , p)} e^{i (p + k_z) n \ell}\\ &=\int \frac{\dif^{d}  k}{(2 \pi)^{d}} \, \int_{-\pi/\ell}^{\pi/ \ell} \frac{\dif p}{2\pi}\, \ell A_\mu( k, p) \avg{J^\mu(- k , -p)} \text{ .}
      \end{split}
    \end{align}  
    Studying the second-order variation to extract the Green's function, we have
    \begin{align}\label{eq:layered_bdy_O2}
      \begin{split}
        &\int \frac{\dif^{d}  k}{(2 \pi)^{d}} \, \int_{-\pi/\ell}^{\pi/ \ell} \frac{\dif p}{2\pi}\, \ell a^\mathsc{s}_\mu( k, p) \delta \avg{J^\mu(- k , -p)}\\
        &= \int \frac{\dif^{d}  k}{(2 \pi)^{d}} \, \int_{-\pi/\ell}^{\pi/ \ell} \frac{\dif p}{2\pi}\, \left(\ell a^\mathsc{s}_\mu( k, p)\right) \frac{G^{\mu\nu}(k)}{\ell} \left(\ell a^\mathsc{s}_\nu(-k, -p)\right)\\
        & = \int \frac{\dif^{d}  k}{(2 \pi)^{d}} \, \int_{-\pi/\ell}^{\pi/ \ell} \frac{\dif p}{2\pi}\, \delta\avg{ J^\mu( k, p)} \left[\ell G^{-1}(k)\right]_{\mu\nu} \delta \avg{J^\nu(k, p)} \text{ ,}
      \end{split}
    \end{align}  
    where $\ell a^\mathsc{s}_\mu( k, p)$ is the $(d + 1)$-dimensional source. In addition, $G(k)$, which is independent of the Bloch momentum $p$, is the Green's function coming from a holographic calculation for a $d$-dimensional boundary theory, so that $G(k)/\ell$ has the dimensions of a $(d + 1)$-dimensional response. When we introduce the double-trace deformation that couples the layers, the total boundary action, second order in fluctuations, then reads
    \begin{align}\label{eq:layered_bdy_green}
      \begin{split}
      &\int \frac{\dif^{d }  k}{(2 \pi)^{d}} \int_{-\pi/\ell}^{\pi/ \ell} \frac{\dif p}{2\pi}\, \delta\avg{J^\mu(- k, -p)} \left[\ell a_\mu( k, p) - \eta_{\mu\nu} \frac{\alpha^2 \ell }{2\abs*{ k}} \frac{\sinh{(\abs*{ k} \ell)}}{\cosh{(\abs*{ k} \ell)} - \cos(p \ell)} \delta \avg{J^\nu( k , p)} \right]\\
      &= \int \frac{\dif^{d }  k}{(2 \pi)^{d}} \int_{-\pi/\ell}^{\pi/ \ell} \frac{\dif p}{2\pi}\, \delta \avg{J^\mu(- k, -p)} \left[\ell G^{-1}_{\mu\nu}(k) - \eta_{\mu\nu} \frac{\alpha^2 \ell}{2\abs*{ k}} \frac{\sinh{(\abs*{ k} \ell)}}{\cosh{(\abs*{ k} \ell)} - \cos(p \ell)} \right] \delta \avg{J^\nu( k , p)}\text{,}
      \end{split}
    \end{align} 
    that gives a $(d + 1)$-dimensional density-density response function of the form
    \begin{align}
      \chi^{00}(k, p) = \frac{G^{00}(k)/\ell}{1 - \frac{\alpha^2}{2\abs*{ k}} \frac{\sinh{(\abs*{ k} \ell)}}{\cosh{(\abs*{ k} \ell)} - \cos(p \ell)} k^2 G^{00}(k)/\bm k^2} \text{ .}
    \end{align}
    Notice that in the limit $\ell \to \infty$ the above boundary action, eq.\ \eqref{eq:layered_bdy_green}, becomes
    \begin{align}
      \int \frac{\dif^{d }  k}{(2 \pi)^{d}} \delta \avg{J^\mu(- k)} \left[G^{-1}_{\mu\nu}(k) - \eta_{\mu\nu} \frac{\alpha^2}{2\abs*{ k}}\right]\delta \avg{J^\nu( k)} \text{ ,}
    \end{align} 
    giving the effective two-dimensional response for a single layer.

\bibliographystyle{JHEP} 

\bibliography{refs}    

\end{document}